\documentclass[prd]{revtex4-1}
\usepackage{float}
\usepackage{graphicx}
\usepackage{caption}
\usepackage{amsmath}
\usepackage{bm}
\usepackage{amssymb}
\usepackage{amsfonts}
\usepackage{color}
\usepackage[normalem]{ulem}

\def\l{{\lambda}}
\def\L{{\Lambda}}
\def\d{{\delta}}
\def\D{{\Delta}}

\def\e{{\epsilon}}

\def\a{{\alpha}}
\def\b{{\beta}}
\def\c{{\chi}}

\def\g{{\gamma}}

\def\p{{\pi}}

\def\m{{\mu}}
\def\n{{\nu}}
\def\r{{\rho}}
\def\s{{\sigma}}
\def\S{{\Sigma}}

\def\x{{\xi}}

\def\({\left(}
\def\){\right)}
\def\[{\left[}
\def\]{\right]}

\newcommand{\pd}{{\partial}}

\newcommand{\tr}{\text{tr}}

\newcommand{\eq}{\text{eq}}

\begin{document}
\title{Spin polarization for massive fermion in a shear flow: complete results at $O(\pd)$}
\author{Ziyue Wang$^1$, Shu Lin$^2$}
\affiliation{
$^1$School of Physics and Optoelectronic Engineering, Beijing University of Technology, Beijing 100021, China\\
$^2$School of Physics and Astronomy, Sun Yat-Sen University, Zhuhai 519082, China
} 

\begin{abstract}
Motivated by the key role of shear induced polarization in understanding the local spin polarization puzzle of $\L$ hyperons in heavy ion collisions, we perform a complete analysis of spin polarization of massive fermion in a quantum electrodynamic plasma with shear flow. Apart from the well-known spin-shear coupling in free theory, we include two more collision dependent contributions: one is a non-dynamical contribution fixed by shifted spin-averaged distribution in steady state; the other is a dynamical contribution following from spin evolution. Despite of the dependencies on collision, we find the dependencies on coupling drop out in the final results. These contributions can lead to significant enhancement of the spin-shear coupling in phenomenologically interesting regime.
\end{abstract}
\maketitle

\section{Introduction}
\label {sec_intro}
The measurement of spin is one of the most challenging and important experimental tasks. The observation of a global spin polarization of $\Lambda$ hyperons \cite{STAR:2017ckg} confirms the interplay between spin and rotation in the relativistic heavy-ion physics \cite{Liang:2004ph,Liang:2004xn}. Such global spin polarization  of final particle can be well described based on a spin-orbit coupling picture \cite{Gao:2007bc,Huang:2011ru,Jiang:2016woz} together with a vorticity in quark-gluon plasma  predicted through model calculations \cite{Deng:2012pc,Pang:2016igs,Xia:2018tes}. However, measurement of $\Lambda$ hyperon local polarization \cite{STAR:2019erd} shows an overall sign difference from theoretical predictions \cite{Becattini:2017gcx,Wei:2018zfb,Fu:2020oxj}. 

Recently it has been realized that shear tensor can also contribute to spin polarization \cite{Liu:2021uhn,Becattini:2021suc,Hidaka:2017auj}.  Phenomenological implementations have shown the right trend toward the measured local polarization results \cite{Fu:2021pok,Becattini:2021iol,Yi:2021ryh,Fu:2022myl,Wu:2022mkr}. The situation is complicated by the fact that shear cannot be consistently described by collisionless theories. It is known that a shear flow necessarily drives the system into a steady state, in which the distribution function has parametrically large deviation from local equilibrium distribution in weakly coupled medium \cite{Arnold:2000dr,Arnold:2003zc}. We illustrated that the large deviation can compensate the smallness in collision term \cite{Lin:2022tma,Lin:2024zik}, giving rise to a collisional contribution to spin polarization, which is parametrically the same as the kinematic contribution obtained in free theory, see also \cite{Fang:2024vds}. 

In the  scenario proposed in \cite{Liang:2004ph}, the spin polarization of $\Lambda$ hyperons inherits the spin of strange quark. The finite mass of strange quark introduces a further twist to the story. Unlike the massless case where spin is locked to the momentum, spin is an independent degree of freedom for massive quark, and is described by an independent kinetic equation. This leads to an extra dynamical contribution to the spin polarization.

All three contributions can be systematically treated in the framework of collisional quantum kinetic theory (QKT). The spin polarization is described by axial component of Wigner function given by \cite{Yang:2020hri,Lin:2021mvw,Hattori:2019ahi}
\begin{equation}
\label{Amu}
\mathcal{A}^{\mu}=2\pi\delta(P^2-m^2)\big(a^\mu f_A+S_{n,m}^{\mu\nu}\mathcal{D}_\nu f\big),
\end{equation}
$a^\mu f_A$ is a dynamical contribution \cite{Hattori:2019ahi,Weickgenannt:2019dks,Gao:2019znl,Liu:2020flb}, which needs to be determined by spin kinetic equation. It can be identified as the Pauli-Lubanski vector, which describes the spin in a given frame specified by the frame vector $n_\m$. $S_{n,m}^{\mu\nu}\mathcal{D}_\nu f$  is the non-dynamical contribution, where $S_{n,m}^{\mu\nu}=\frac{\epsilon^{\mu\nu\alpha\beta}P_\alpha n_\beta}{2(P\cdot n+m)}$ and $\mathcal{D}_\nu f=\partial_\nu f+\Sigma_{V\nu}^>f-\Sigma_{V\nu}^<\bar{f}$. $f$ and $\bar{f}=1-f$ are the spin-averaged distribution function and the Pauli blocking factor. $\S^{\gtrless}_{V\n}=\frac{1}{4}\tr\[\g_\n\S^{\gtrless}\]$ corresponds to vector component of greater/lesser self-energy accounting for the collisional contribution. The $\pd f$ and $\S f$ terms are recognized as the magnetization current and the displacement current, corresponding to kinematic and collisional contributions to spin polarization respectively \cite{Lin:2024zik}. In \cite{Lin:2022tma} we determined the collisional contribution for massive fermion in a shear flow, leaving the dynamical contribution undetermined.

In this paper, we fill the gap by determining the dynamical contribution, thus providing complete results for spin polarization at $O(\pd)$. We shall treat the less abundant strange quark as probe to the medium consisting of light quarks and gluons. Since the QKT framework for QCD is not available for now, we study a simplified version of this problem: probe massive fermion in massless QED plasma with a shear flow. In the limit of large fermion mass, Compton scattering and pair annihilation are suppressed. The collisional and dynamical contributions to probe fermion come entirely from Coulomb scattering with medium fermions.
The collisional contribution has been studied in our previous work \cite{Lin:2022tma}, where we assumed the probe fermion in local equilibrium while the medium fermions reach steady state. We found the collisional contribution suppresses the kinematic contribution. In this work, we will consider both probe and medium fermions in steady state. The dynamical contribution will be determined for the first time. We shall show the spin kinetic equation leads to a detailed balance condition with vanishing collision term in the long time limit, which allows us to fix the dynamical contribution.

This paper is organized as follows, in Sec.\ref{sec_selfenergy} we collect and calculate the non-dynamical contribution to the massive probe fermion. In Sec.\ref{sec_frame} we review the frame independence of the axial-vector component of Wigner function $\mathcal{A}_\mu$ and use it to determine $\mathcal{A}_\mu$ for massless fermion. We further generalize the procedure to determine the leading mass correction to $\mathcal{A}_\mu$ for massive fermions. In Sec.\ref{sec_staticsolu}, we derive the collision term of spin kinetic equation. The equation is analyzed in the long time limit, in which we obtain the static solution of $\mathcal{A}_\mu$ and the dynamical term $a_\mu f_A$ from the vanishing of the collision term. In Sec.\ref{s5} we provide conclusion and outlook. Calculation details are presented in Appendix.\ref{details}.

\section{Self-energy correction}
\label{sec_selfenergy}
We start with some general analysis based on symmetries. For the spin polarization induced by hydrodynamic gradients, it is naturally to count it as $\mathcal{A}_\mu\sim\mathcal{O}(\partial)$. In the fluid with shear tensor only, $\mathcal{A}_\mu$ can be decomposed into 
\begin{eqnarray}
\label{decomposeA}
\mathcal{A}_\mu=2\pi\delta(P^2-m^2)\left(u_\mu I^P_{\nu\xi}N_{1p}+\hat{P}_{\perp\mu}I^P_{\nu\xi}N_{2p}+J^P_{\mu,\nu\xi}N_{3p}+L^P_{\mu,\nu\xi}N_{4p}\right)f'_p\sigma^{\nu\xi}.
\end{eqnarray}
We have used the abbreviation $f'_p=\partial_{\beta u\cdot P}f_p(\b u\cdot P)$ and $f_p(\b u\cdot P)\equiv f_V^{eq}(\b u\cdot P)=\(\exp(\b u\cdot P)+1\)^{-1}$, with $\b$ and $u^\m$ being the inverse temperature and fluid velocity. The projectors $I^P_{\nu\xi}$, $J^P_{\mu,\nu\xi}$ and $L^P_{\mu,\nu\xi}$ are defined as
\begin{eqnarray}
\label{projector}
I_{\xi\nu}^P&=&\hat{P}_{\perp\xi}\hat{P}_{\perp\nu}+\frac{1}{3}\Delta_{\xi\nu},\nonumber\\
J_{\mu,\xi\nu}^P&=&-\hat{P}_{\perp\xi}\Delta_{\mu\nu}-\hat{P}_{\perp\nu}\Delta_{\mu\xi}+\frac{2}{3}\hat{P}_{\perp\mu}\Delta_{\xi\nu},\nonumber\\
L_{\mu,\xi\nu}^P&=&\frac{1}{2}(\epsilon_{\mu\xi\lambda\sigma}u^\lambda\hat{P}_{\perp}^\sigma\hat{P}_{\perp\nu}+\epsilon_{\mu\nu\lambda\sigma}u^\lambda\hat{P}_{\perp}^\sigma\hat{P}_{\perp\xi}).
\end{eqnarray}
As has been discussed in \cite{Liu:2021uhn}, $\mathcal{A}^\mu$ transforms under time reversal as $\mathcal{A}^\mu\rightarrow ( \mathcal{A}^0,- \mathcal{A}^i)$ and transforms under parity reversal as $\mathcal{A}^\mu\rightarrow (-\mathcal{A}^0,\mathcal{A}^i)$. In the absence of parity odd parameter such as $\m_5$, $N_{1p}$ through $N_{4p}$ can only be parity even functions. In follows that only term with $L_{\mu,\nu\xi}\sigma^{\nu\xi}$ in (\ref{decomposeA}) has the same behavior under time and parity reversal. Thus $\mathcal{A}_\mu$ can be written as 
\begin{eqnarray}
\label{defineNP}
\mathcal{A}_\mu=2\pi \delta(P^2-m^2)\frac{\epsilon_{\mu\nu\alpha\beta}P_\perp^\alpha u^\beta}{2P\cdot u}P_{\perp\xi}\sigma^{\nu\xi}f'_p N_P.
\end{eqnarray}
On the other hand, in a collisional QKT \cite{Yang:2020hri,Lin:2021mvw,Hattori:2019ahi}, the axial component of Wigner function for fermion in the local rest frame of the fluid $n_\mu=u_\mu$ is given by \eqref{Amu}. $\mathcal{A}_\mu$ contains the dynamical part $a_\mu f_A$ as well as non-dynamical part $S_{u,m}^{\mu\nu}\mathcal{D}_\nu f$, with  $\mathcal{D}_\nu f=\partial_\nu f+\Sigma_{V\nu}^>f-\Sigma_{V\nu}^<\bar{f}$. With the same logic, we can parameterize the dynamical, kinematic and collisional contributions separately as
\begin{eqnarray}
\label{defineNi}
\mathcal{A}_\mu^a&=&2\pi \delta(P^2-m^2) a_\mu f_A~=~2\pi \delta(P^2-m^2)\frac{\epsilon_{\mu\nu\alpha\beta}P_\perp^\alpha u^\beta}{2P\cdot u}P_{\perp\xi}\sigma^{\nu\xi}f'_p {N_a},\nonumber\\
\mathcal{A}_\mu^\partial&=&2\pi \delta(P^2-m^2) S^{m,u}_{\mu\nu}\partial^\nu f_p~=~2\pi \delta(P^2-m^2)\frac{\epsilon_{\mu\nu\alpha\beta}P_\perp^\alpha u^\beta}{2P\cdot u}P_{\perp\xi}\sigma^{\nu\xi}f'_p {N_\partial},\nonumber\\
\mathcal{A}_\mu^\Sigma&=&2\pi \delta(P^2-m^2) S^{m,u}_{\mu\nu}\widehat{\Sigma_{V}^{\nu}f_p}~=~2\pi \delta(P^2-m^2)\frac{\epsilon_{\mu\nu\alpha\beta}P_\perp^\alpha u^\beta}{2P\cdot u}P_{\perp\xi}\sigma^{\nu\xi}f'_p {N_\Sigma},
\end{eqnarray}
The quantities $N_\pd$ and $N_\S$ can be further related by scalar kinetic equation. Using the rotational symmetry, $\Sigma_{V\nu}^>f-\Sigma_{V\nu}^<\bar{f}$ in the shear flow can be decomposed as 
\begin{eqnarray}
\label{self-energy-decompose}
\Sigma_{V\nu}^>f-\Sigma_{V\nu}^<\bar{f}&=&(u_\nu I_{\alpha\beta}T_{1p}+\hat{P}_{\perp\nu}I_{\alpha\beta}T_{2p}+J_{\nu,\alpha\beta}T_{3p})f'_p\sigma_{\alpha\beta}.
\end{eqnarray}
Thus $\mathcal{D}_\nu f$ becomes
\begin{eqnarray}
\label{decompositionDf}
&&\mathcal{D}_\nu f=P^\alpha\sigma_{\alpha\nu}f'_p+(u_\nu I^{\alpha\beta}T_{1p}+\hat{P}_{\perp\nu}I^{\alpha\beta}T_{2p}+J_{\nu}^{\alpha\beta}T_{3p})f'_p\sigma_{\alpha\beta}.
\end{eqnarray}
The scalar kinetic equation $P^\nu\mathcal{D}_\nu f=0$ \cite{Yang:2020hri,Lin:2021mvw} gives a constraint between the above parameters,
\begin{eqnarray}
\label{constraintT1}
&&T_{1p}=\frac{p^2}{P\cdot u}\Big(\frac{T_{2p}}{p}+\frac{2T_{3p}}{p}-1\Big).
\end{eqnarray}
With the decomposition above, the dynamical part, the kinematic part and the collisional part have the following decomposition, 
and the axial-vector component as the sum of the above three parts $\mathcal{A}_\mu=\mathcal{A}_\mu^a+\mathcal{A}_\mu^\partial+\mathcal{A}_\mu^\Sigma$. Comparing \eqref{defineNP} with \eqref{defineNi} one can directly find $N_p$ is the sum of three parts
\begin{eqnarray}
\label{NPdecompose}
N_P=N_a+N_\partial+N_\Sigma=N_a+\frac{P\cdot u}{P\cdot u+m}\Big(1-\frac{2T_{3p}}{p}\Big),
\end{eqnarray}
where in the last equality we have used $N_\pd=\frac{P\cdot u}{P\cdot u+m}$ and $N_\S=-\frac{P\cdot u}{P\cdot u+m}\frac{T_3}{p}$ following from \eqref{defineNi} and \eqref{self-energy-decompose} respectively.


Before moving on to the detailed calculation, we first recall the results in related studies. Results obtained in linear response theory \cite{Liu:2021uhn} and statistical field theory \cite{Becattini:2021suc} correspond $N_P=1$. Chiral kinetic theory also gives $N_P=1$ in the massless limit \cite{Hidaka:2017auj}. Its extrapolation to massive case corresponds to $N_\pd$ \cite{Yi:2021ryh}. All three results do not contain collision dependent contributions, which are generically present in $N_a$ and $N_\S$. 
As we are going to see, $\mathcal{A}_\mu$ for massive spin carriers with collisional effect determined in a static state is much more complicated.  

We first work out the collisional contribution for probe fermion in the massless QED plasma with a steady shear flow with both probe and medium fermions (and also photons) in steady state. We start by collecting the result in our previous work \cite{Lin:2024zik}. Consider a neutral QED plasma with $N_f$ flavor of massless fermions in a shear flow. To the leading logarithmic order, only elastic scatterings among fermions and photons are relevant. The presence of shear flow leads to redistribution of fermions and photons, which has been solved in Ref.\cite{Arnold:2002zm,Arnold:2000dr,Arnold:2003zc}. The deviation from local equilibrium distribution can be parametrized as $\delta f_p=f_p(1-f_p)I^p_{ij}\sigma_{ij}\chi_p$ with the thermal shear $\sigma_{ij}=\frac{1}{2}(\partial_i\beta_j+\partial_j\beta_i)-\frac{1}{3}\delta_{ij}\partial\cdot\beta$ and $I^p_{ij}=\hat{p}_i\hat{p}_j-\frac{1}{3}\delta_{ij}$ being a symmetric traceless tensor defined with 3-momentum $p$. Approximate analytic expressions have been obtained in \cite{Lin:2022tma}, the redistribution for massless medium fermion is characterized by $\delta f_p$ with $\chi_p$ given by
\begin{eqnarray}
\label{chi_medium}
\chi_p^{\text{med}}=\frac{(2\pi)^3C_f}{e^4\ln e^{-1}}\beta^2p^2,
\end{eqnarray}
with $C_f=\frac{3(1+2N_f)}{4\pi^2N_f^2}$. This redistribution leads to collisional contribution through the self-energy correction. The displacement contribution of the massless medium fermion includes Coulomb, Compton and annihilation processes \cite{Lin:2024zik}. We quote the result in \cite{Lin:2024zik} for the spin polarization of massless fermion in a shear flow with vanishing dynamical contribution
\begin{eqnarray}
\mathcal{A}_\mu&=&2\pi \delta(P^2) S_{\mu\nu}\mathcal{D}^\nu f_p\nonumber\\
&=&-2\pi \delta(P^2)\frac{\epsilon_{\mu\nu\alpha\beta}P_\perp^\alpha u^\beta}{2P\cdot u}P_{\perp\xi}\sigma^{\nu\xi}\Big(1-\frac{2T_{3p}^{\text{med}}}{p}\Big)f^{\eq}_p(1-f^\eq_p).
\end{eqnarray}
The ratio $\frac{2T_{3}}{p}$ delineates the importance of collisional contribution in comparison to the well investigated kinematic contribution. $T_3$ is the coefficient contributed from the self-energy correction \cite{Lin:2024zik}. 
\begin{eqnarray}
\label{T3med}
T_{3p}^{\text{med}}&=&\frac{C_f N_f \left(9 T \zeta (3) (5 F p-4 T)+\pi ^2 p (4 T-F p)\right)}{6 p}\nonumber\\
&&+\frac{(C_f-C_b)\pi ^4 T^2 }{48 p} \frac{1+f^\eq_{\gamma, p}}{1-f^\eq_p}.
\end{eqnarray}
$T_3$ splits into contribution from Coulomb scattering (first line) and from Compton and annihilation (second line), with $F=1-2f_p^\eq$, $f^\eq$ and $f_\g^\eq$ are the distribution functions for fermions and photons at local equilibrium respectively, and $(C_b-C_f)=\frac{2}{\pi^2N_f}$. 

\begin{figure}[H]\centering
\includegraphics[width=0.35\textwidth]{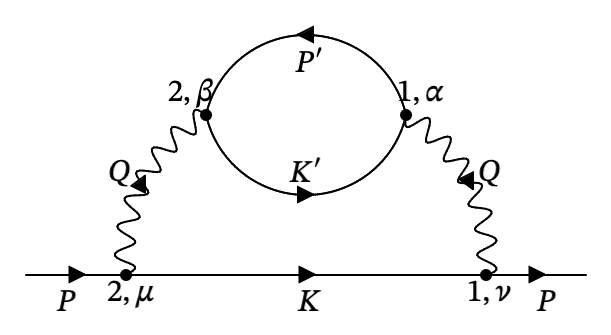}
\caption{Two-loop diagram for fermion self-energy containing propagator corrections \cite{Lin:2021mvw, Wang:2022yli}. The unprimed momenta and primed momenta correspond to probe and medium fermions respectively.}
\label{fig1}
\end{figure}

Next, we work out the self-energy part $\mathcal{A}_\mu^\Sigma$ in the spin polarization of the probe fermion. We assume the probe fermion also reaches a steady state. At the leading logarithmic order we consider, the Compton scattering and pair annihilation are suppressed as long as $m\gg eT$. This is because the Compton scattering and pair annihilation involving the probe fermion are screened by the its mass, giving rise to an infrared enhancement factor $\ln\frac{T}{m}$, which is much less than the counterpart $\ln\frac{T}{eT}$ from Coulomb scattering. Therefore, we can obtain the redistribution of probe fermion by consider Coulomb scattering with medium fermions only. The self-energy diagram corresponding to Coulomb scattering is depicted in Fig.~\ref{fig1}. 
The relevant scalar kinetic equation for the massive probe fermion reads
\begin{eqnarray}
\label{Boltzmann}
P\cdot \partial f_p=-4e^4(2\pi^3)\int_{K,P',K'}(2\pi)^4(P-K-K'+P')\delta(K^2-m^2)\delta(P'^2)\delta(K'^2)|{\mathcal M}|_{\text{Coul}}^2\Big(f_pf_{p'}\bar{f}_{k}\bar{f}_{k}-\bar{f}_p\bar{f}_{p'}{f}_{k}{f}_{k'}\Big),
\end{eqnarray}
where $|{\mathcal M}|_{\text{Coul}}^2$ is the amplitude square for Coulomb scattering with overall coupling constants factored out 
\begin{eqnarray}
|{\mathcal M}|_{\text{Coul}}^2=\frac{2N_f  \cdot 2(K\cdot P'P\cdot K'+P\cdot P' K\cdot K'-m^2P'\cdot K')}{(Q^2)^2}.
\end{eqnarray}
The unprimed momenta and primed momenta correspond to probe and medium fermions respectively. $Q=P-K=K'-P'$ is the small momentum transfer. $2N_f$ in the numerator comes from the $N_f$ flavor fermion and anti-fermion in the loop. The redistribution of medium fermions is already known. The counterpart of probe fermion can be parameterized as $\d f_{p}=f_{p}(1-f_{p})I_{ij}^{p}\s_{ij}\c_p^{\text{prob}}$. 
After linearizing the Boltzmann equation and performing the phase space integral, \eqref{Boltzmann} is then turned into a differential equation of $\chi_p$
\begin{eqnarray}\label{diff_chi_probe}
\frac{2}{3}p^2f_p(1-f_p)&=&\frac{e^4\ln e^{-1}}{(2\pi)^4 }\frac{4 \pi ^3 T^3 N_f}{3p \tilde{v}^4}\Big\{
\Big(\tilde{v}(3 \tilde{v}^2-1)+(\tilde{v}^2-1)^2 \eta_p\Big)\chi_p^\text{prob}
\nonumber\\
&&+\frac{p \tilde{v}^2}{3 T} \left(F p \left(\tilde{v}^2-1\right) \eta _p+F p \tilde{v}-2 T \tilde{v}^4\right){\chi_p^\text{prob}}'
-\frac{1}{3} p^2 \tilde{v}^2 ((\tilde{v}^2-1) \eta_p+\tilde{v}){\chi_p^\text{prob}}''\nonumber\\
&&+\frac{72 \pi p \zeta (3)C_f}{T^2}\left(\tilde{v}^2-1\right) \left(F p \left(\tilde{v}^2-3\right) \eta _p+3 F p \tilde{v}+4 T \tilde{v}^4\right)\Big\}f_p(1-f_p),
\end{eqnarray}
where $F=1-2f_p$ and $\tilde{v}=p/p_0$, the rapidity $\eta_p\equiv\text{arctanh}\,\tilde{v}\equiv2^{-1}\ln[(p_0+p)/(p_0-p)]$, with $p_0=(p^2+m^2)^{1/2}$. The redistribution of the probe fermion in the shear flow can be characterized with $\chi_p^{\text{prob}}$. We solve \eqref{diff_chi_probe} in the phenomenological interesting regime $p\gg T$. An approximate analytical solution can be found by keeping only the leading powers of $p$ in \eqref{diff_chi_probe}, which singles out $\c_p^{\text{prob}'}$ terms on the RHS. It follows then 
\begin{eqnarray}\label{chi_probe}
\chi_p^{\text{prob}}=\frac{(2\pi)^3\,C_f^{\text{prob}}}{e^4\ln e^{-1}}\beta^2p^2.
\end{eqnarray}
In comparison to \eqref{chi_medium}, $C_f^{\text{prob}}$ of the probe fermion is a function of momentum and mass
\begin{eqnarray}
C_f^{\text{prob}}=\frac{3}{2\pi ^2N_f}\cdot\frac{\tilde{v}^2}{\tilde{v}+(\tilde{v}^2-1) \text{arctanh}\,\tilde{v}},
\end{eqnarray}
with $\tilde{v}=p/p_0$. In the massless limit, namely as $\tilde{v}\rightarrow 1$, the above coefficient has the limit $C_f^{\text{prob}}\rightarrow \frac{3}{2\pi ^2N_f}$. This is different from the corresponding coefficient $C_f=\frac{3(1+2N_f)}{4\pi^2N_f^2}$ of the massless medium fermion since only Coulomb scattering is considered for massive fermion. 

\begin{figure}[H]\centering
\includegraphics[width=0.45\textwidth]{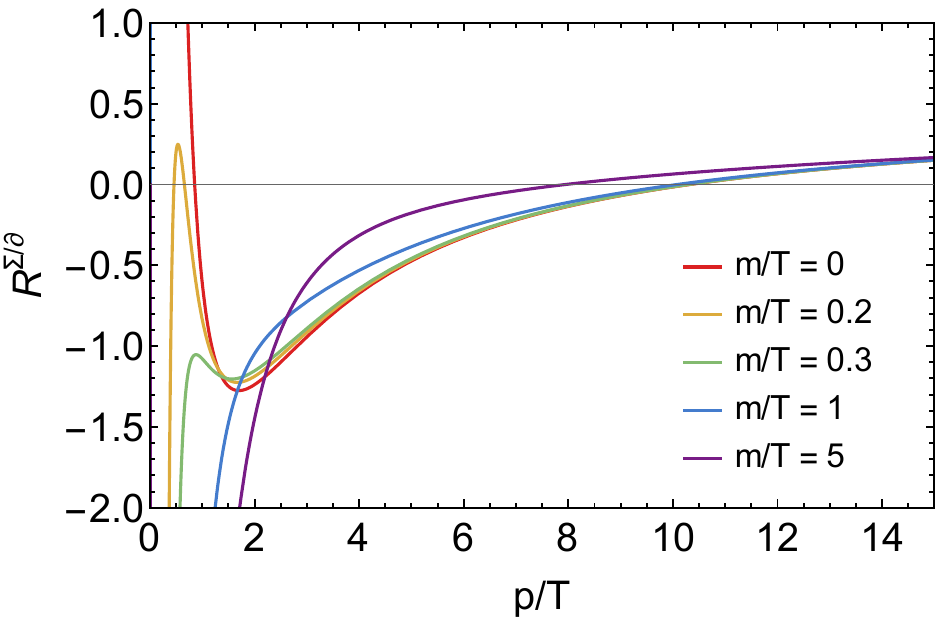}
\caption{Ratio $R^{\Sigma/\partial}\equiv N_\Sigma/N_\partial=-2T_{3p}^{\text{prob}}/p$ versus $p/T$ for different masses for $N_f=2$. }
\label{Nratio1}
\end{figure}

With the redistribution of the medium and probe fermion determined above, we can work out $\Sigma_{V\nu}^>f-\Sigma_{V\nu}^<\bar{f}\equiv\widehat{\Sigma_{V\mu}f}$ for the probe fermion. 
\begin{eqnarray}
\label{prob-self-energy}
\widehat{\Sigma_{V\mu}f_p}
&=&4e^4(2\pi)^3\int_{K',P',K}(2\pi)^4\delta(P-K+P'-K')\delta(K^2-m^2)\delta(P'^2)\delta(K'^2)\epsilon(k_0)\epsilon(k'_0)\epsilon(p'_0)\nonumber\\
&&\qquad\qquad\qquad\quad\times M_\mu^\text{Coul}\big({f}_{p'} {f}_{p}\bar{f}_{k}\bar{f}_{k'} -\bar{f}_{p}\bar{f}_{p'}{f}_{k}{f}_{k'}\big)\nonumber\\
&=&4e^4(2\pi)^3\int_{K',P',K}(2\pi)^4\delta(P-K+P'-K')\delta(K^2-m^2)\delta(P'^2)\delta(K'^2)\epsilon(k_0)\epsilon(k'_0)\epsilon(p'_0)\nonumber\\
&&\qquad\qquad\qquad\quad\times M_\mu^\text{Coul}\Big[I_{\alpha\beta}^{\hat{P}'}\chi_{p'}^{\text{med}}-I_{\alpha\beta}^{\hat{K}'}\chi_{k'}^{\text{med}}+I_{\alpha\beta}^{\hat{P}}\chi_{p}^{\text{prob}}-I_{\alpha\beta}^{\hat{K}}\chi_{k}^{\text{prob}}\Big]f_p f_{p'}\bar{f}_k\bar{f}_{k'}\sigma^{\alpha\beta}\nonumber\\
&=&T^{\text{prob}}_{\mu,\alpha\beta}(P)\sigma^{\alpha\beta},
\end{eqnarray}
with 
\begin{eqnarray}
M_\mu^\text{Coul}=\frac{2N_f\cdot 2\left(P'_\mu K\cdot K'+K'_\mu K\cdot P'\right)}{(Q^2)^2}.
\end{eqnarray}
Note that the general analysis from \eqref{self-energy-decompose} through \eqref{constraintT1} applies to $\widehat{\Sigma_{V\mu}f}$, so that we can write
\begin{align}
\widehat{\Sigma_{V\mu}f}=(u_\nu I_{\alpha\beta}T_{1p}^{\text{prob}}+\hat{P}_{\perp\nu}I_{\alpha\beta}T_{2p}^{\text{prob}}+J_{\nu,\alpha\beta}T_{3p}^{\text{prob}})f'_p\sigma_{\alpha\beta},
\end{align}
with
\begin{eqnarray}
\label{T2probe}
T_{2p}^{\text{prob}}&=&
\frac{6C_f N_f   T \zeta(3)}{p \tilde{v}^4}\Big[\left(9 \tilde{v}^4-7\tilde{v}^2+15\right)\tilde{v}^2T  +\left(17 \tilde{v}^2-30\right)p \tilde{v} F+3\left(\left(\tilde{v}^4-9 \tilde{v}^2+10\right)p F +\left(3 \tilde{v}^2-5\right)\tilde{v} T \right) \eta_p \Big]\nonumber\\
&&
-\frac{C_f^{\text{prob}}N_f  \pi ^2}{3 \tilde{v}^6}\Big[\left(4 \tilde{v}^6+2 \tilde{v}^4+9 \tilde{v}^2+3\right)\tilde{v} T -\left(4
   \tilde{v}^2+3\right)p \tilde{v}^2 F+\left(\left(-2 \tilde{v}^4+3 \tilde{v}^2+3\right)p \tilde{v} F + \left(2 \tilde{v}^6+\tilde{v}^4-8 \tilde{v}^2-3\right)T\right)\eta _p \Big],\\
\label{T3probe}
T_{3p}^{\text{prob}}&=&\frac{3T C_f N_f  \zeta (3)}{2 p^3 \tilde{v}^3} \Big[2\left( -3 \tilde{v}^4+7\tilde{v}^2-6\right)p^2\tilde{v}  T+\left(24-19 \tilde{v}^2\right)p^3 F+3 m^2 \tilde{v} \left(\left(\tilde{v}^2-8\right)p F +4 T \tilde{v}\right) \eta _p \Big]\nonumber\\
&&+\frac{C_f^{\text{prob}}N_f  \pi ^2 }{6 p^2 \tilde{v}^5}\Big[2 \left( -4 \tilde{v}^4+9 \tilde{v}^2 -3\right)p^2 T+\left(2 \tilde{v}^2-3\right)p^3 \tilde{v} F +m^2 \tilde{v} \left(3 p \tilde{v}F + \left(6-14 \tilde{v}^2\right)T\right)\eta _p\Big].
\end{eqnarray}
In massless limit, $T_{3p}^{\text{prob}}$ becomes
\begin{eqnarray}
\label{T3massless}
T_{3p}^{\text{prob},m=0}&=&\frac{3C_f N_f T \zeta (3) (5 F p-4 T)}{2 p}+\frac{C_f^{\text{prob}} N_f\pi ^2 (4 T-F p)}{6}.
\end{eqnarray}
The ratio $R^{\Sigma/\partial}\equiv N_\Sigma/N_\partial=-2T_{3p}^{\text{prob}}/p$ delineates the importance of collisional contribution $A_\mu^{\Sigma}$ in comparison to the kinematic contribution $A_\mu^{\partial}$. As has been discussed in \cite{Lin:2024zik}, the sign of this ratio depends on $p$. The ratio $R^{\Sigma/\partial}$ versus $p/T$ for different masses is presented in Fig.\ref{Nratio1}. The ratio for $p\lesssim T$ is unreliable because of the use of approximate solution \eqref{chi_probe}.
Note that in the limit $p\gg T$, the ratio $R^{\Sigma/\partial}$ still depends the mass of the probe fermion. In the relativistic limit $p\gg m$, 
\begin{eqnarray}
R^{\Sigma/\partial}|_{p\gg m,p\gg T}\rightarrow \frac{1}{2}.
\end{eqnarray}
In the non-relativistic limit  $m\gg p$, 
\begin{eqnarray}
R^{\Sigma/\partial}|_{m\gg p}\rightarrow -\frac{18 C_f N_f \zeta (3) T^2 }{p^2}-\frac{2 m T}{5 p^2}.
\end{eqnarray}

\section{Frame dependence}
\label{sec_frame}
In this section, we constrain the dynamical part from the frame independence of $\mathcal{A}_\mu$. We start our analysis by collecting the known result in \cite{Chen:2015gta}, see also \cite{Sheng:2022ssd}. It has long been realized that Lorentz covariance and the requirement of conservation of angular momentum together generate a frame-dependent particle position, which is a tunneling-like motion of the particle known as the side-jump. For a massless fermion with helicity $\lambda$, the Lorentz covariant current in presence of a 2-by-2 collision $PK\rightarrow P'K'$ in a frame specified by vector $n^\m$ is given by
\begin{equation}
J_\lambda^\mu=P^\mu f_\lambda+\lambda S^{\mu\nu}_n\partial_\nu f_\lambda+\lambda\int_{KK'P'} C^\lambda \bar{\Delta}^\mu,
\end{equation}
with $\lambda=\pm$ corresponding to right/left-handed particles. $\lambda S^{\mu\nu}_n=\lambda\frac{\epsilon^{\mu\nu\alpha\beta}P_\alpha n_\beta}{2P\cdot n}$ is the spin tensor for fermion with helicity $\lambda$ in frame $n$, $\lambda\bar{\Delta}^\mu\equiv\lambda\Delta_{\bar{n}n}^\mu=\lambda\frac{\epsilon^{\mu\alpha\beta\gamma}P_\alpha \bar{n}_\beta n_\gamma}{2(P\cdot n)(P\cdot \bar{n})}$ is the side-jump of the particle between the "no-jump frame" $\bar{n}=(P+K)/\sqrt{s}$ and the arbitrary frame $n^\m$. $\int_{KK'P'}$ is the phase space integral and $C^\lambda$ is the collision kernel with the following explicit form\footnote{We have dropped the symmetry factor $\frac{1}{2!}$ irrelevant for our case with probe fermion.}.
  \begin{align}\label{coll_kern}
    C^\l=|{\cal M}|^2(2\p)^4\d(P+K-P'-K')\big[f^\l_{\bar{n}}(P')f^\l_{\bar{n}}(K')(1-f^\l_{\bar{n}}(P))(1-f^\l_{\bar{n}}(K))-(P,K\leftrightarrow P',K')\big],
  \end{align}
with $f_{\bar{n}}$ being the distribution function in the no-jump frame.
The axial-vector current can be constructed from $J_\lambda^\mu$ as 
\begin{equation}
\label{Amu1}
\mathcal{A}^\mu=J_+^\mu-J_-^\mu=P^\mu f_A+ S^{\mu\nu}_n\partial_\nu f_V+\int_{KK'P'} C[f_V] \bar{\Delta}^\mu.
\end{equation}
With a little bit of algebra, one can translate (\ref{Amu1}) into the corresponding expression in \cite{Hidaka:2016yjf}
\begin{equation}
\label{Amu2}
\mathcal{A}^{\mu}=2\pi\delta(P^2)\Big(P^\mu f_A+S_{n}^{\mu\nu}\mathcal{D}_\nu f_V\Big),
\end{equation}
The generalization from massless to massive case \eqref{Amu} is minimal \cite{Hattori:2019ahi}. By construction, $a^\mu f_A$ is identified as the Pauli-Lubanski vector which describes the spin in a given frame. It is a dynamical variable which needs to be determined by the kinetic theory. On the other hand, $S_{n,m}^{\mu\nu}\partial_\nu f_V$ can be recognized as the magnetization current and $S_{n,m}^{\mu\nu}\widehat{\Sigma_{V\nu} f_V}$ as the displacement current.

Since ${\cal A}^\m$ is frame independent, it requires that the frame dependences in dynamical and non-dynamical part to cancel. As we shall show below this requirement can fully fix $f_A$ in the massless case and can fix the leading mass correction to $a^\m f_A$ in the massive case. We begin with the massless case, in which the requirement leads to
\begin{equation}
P^\m(f'_A-f_A)=\(S^{\m\n}_n-S^{\m\n}_{n'}\)\mathcal{D}_\nu f_V=\(P^\n\D_{nn'}^\m-P^\m\D_{nn'}^\n\)\mathcal{D}_\nu f_V,
\end{equation}
where $f'_A$ and $f_A$ are the axial-charge in frame $n'$ and $n$ respectively. The frame dependence in $f_V$ occurs only at higher order in gradient, so we may take $f_V$ in any frame. Using the kinetic equation $P\cdot{\cal D}f_V=0$, we easily find
\begin{align}
  f'_A-f_A=-\D_{nn'}^\n\mathcal{D}_\nu f_V.
\end{align}
We proceed by writing down the general frame independent decomposition of $\mathcal{D}_\nu f_V$ with the shear tensor $\s_{\a\b}$ as
\begin{align}
  -\mathcal{D}_\nu f_V=P^\a P^\b\s_{\a\b}\(A_1P_\n+A_2u_\n\)+A_3P^\m\s_{\n\m}.
\end{align}
The term proportional to $P_\n$ vanishes upon contraction with $\D_{nn'}^\n$. The coefficients $A_2$ and $A_3$ are related by the kinetic eqution $P\cdot{\cal D}f_V=0$ as $A_2=-\frac{1}{P\cdot u}A_3$. Thus we have
\begin{align}
  f_A'-f_A=\frac{\e^{\m\n\a\b}P_\n n_\a n_\b'}{2(P\cdot n)(P\cdot n')}\(-\frac{P^\x P^\r\s_{\x\r}}{P\cdot u}u_\m+p^\x\s_{\m\x}\)A_3.
\end{align}
Using the following contracted Schouten identity
\begin{align}
  \(\e^{\m\n\a\b}P^\x+\e^{\n\a\b\x}P^\m+\e^{\a\b\x\m}P^\n+\e^{\b\x\m\n}P^\a+\e^{\x\m\n\a}P^\b\)P_\n n_\a n_\b'\frac{P^\r\s_{\x\r}u^\m}{(P\cdot n)(P\cdot n')(P\cdot u)}=0,
\end{align}
and noting the middle term vanishes by the on-shell condition $P^2=0$, we obtain
\begin{align}
  f_A'-f_A=\(\frac{\e^{\m\n\a\b}u_\n P_\a n_\b'P^\x\s_{\m\x}}{2P\cdot n'}-(n'\to n)\)\frac{A_3}{P\cdot u}.
\end{align}
It is solved by $f_A=\frac{\e^{\m\n\a\b}u_\n P_\a n_\b P^\x\s_{\m\x}}{2P\cdot n}\frac{A_3}{P\cdot u}$. In particular, when we take $n^\m=u^\m$, $f_A=0$.

For massive fermion, we use a frame independent vector $N^\m$ to parametrize $a^\m f_A$ as 
\begin{align}
\label{Nmu}
a^\m f_A+\frac{\e^{\m\n\a\b}P_\a n_\b{\cal D}_\n f_V}{2(P\cdot n+m)}\equiv N^\m.
\end{align}
$N^\m$ should be chosen such that $f_A$ has the correct massless limit and $a^\m$ satisfies the constraint $P\cdot a=P^2-m^2=0$ \cite{Hattori:2019ahi}. Recall that for $n^\m=u^\m$, we have the following explicit expression for $N^\m$ in the massless limit.
\begin{align}
N^\m(m=0)&=\frac{\e^{\m\n\a\b}P_\a u_\b{\cal D}_\n f_V}{2P\cdot u} \nonumber\\
&=\frac{\e^{\m\n\a\b}P_\a u_\b}{2P\cdot u}P_{\perp\x}\s^{\n\x}f_V'\(1-\frac{2T_{3p}(m=0)}{p}\).
\end{align}
In small mass limit, we choose the corresponding $N^\m$ to have the same expression as above except with $T_{3p}(m=0)$ replaced by the massive counterpart. Inserting the decomposition \eqref{decompositionDf} back to \eqref{Nmu}, we then have
\begin{eqnarray}
a^\mu f_A&=&-\frac{\epsilon^{\mu\nu\alpha\beta}P_\alpha u_\beta}{2P\cdot u}\Big(\frac{2T_{3p}}{p}-1\Big)f'_P{P}^{\xi}\sigma_{\xi\nu}\nonumber\\
&&-\frac{\epsilon^{\mu\nu\alpha\beta}P_\alpha n_\beta u_\nu }{2(P\cdot n+m)P\cdot u}\Big(\frac{T_{2p}}{p}+\frac{2T_{3p}}{p}-1\Big)f'_PP^\xi P^\sigma \sigma_{\xi\sigma}
+\frac{\epsilon^{\mu\nu\alpha\beta}P_\alpha n_\beta}{2(P\cdot n+m)}\Big(\frac{2T_{3p}}{p}-1\Big)f'_PP^{\xi}\sigma_{\xi\nu}.
\end{eqnarray}
Using the Schouten identity, the above dynamical term $a^\mu f_A$ can be rewritten as 
\begin{eqnarray}
a^\mu f_A&=&\Big(P^\mu\frac{\epsilon^{\nu\alpha\beta\rho}P_\alpha n_\beta u_\nu }{2(P\cdot n+m)P\cdot u}
+m^2\frac{\epsilon^{\mu\nu\alpha\rho}n_\alpha u_\nu }{2(P\cdot n+m)P\cdot u}
+m\frac{\epsilon^{\mu\nu\alpha\rho}P_\alpha u_\nu }{2(P\cdot n+m)P\cdot u}\Big)\Big(\frac{2T_{3p}}{p}-1\Big)f'_PP^\xi \sigma_{\xi\rho}
\nonumber\\
&&-m^2\frac{\epsilon^{\mu\nu\alpha\beta}P_\alpha n_\beta u_\nu }{2(P\cdot n+m)P\cdot u}\frac{T_{2p}}{p}f'_PP^\xi P^\sigma\sigma_{\xi\sigma}.
\end{eqnarray}
It is easy to find from the above expression $P^\mu a_\mu=0$ is satisfied. 
Taking $n^\mu=u^\mu$, we have 
\begin{eqnarray}
\label{Nalinearm}
a^\mu f_A&=&m\Big(1-\frac{2T_{3p}(m=0)}{p}\Big)\frac{\epsilon^{\mu\nu\alpha\beta}P_\alpha u_\beta }{2(P\cdot u)^2}f'_PP^\xi \sigma_{\xi\nu}+\mathcal{O}(m^2).
\end{eqnarray}
Now we comment on the uniqueness of the ansatz: the structure of $N^\m$ is fixed by Lorentz covariance and parity. The normalization is subject to possible mass correction as $1-\frac{2T_{3p}(m=0)}{p}\to 1-\frac{2T_{3p}(m=0)}{p}+O(m^2)$. However, the $O(m^2)$ correction will not affect our result \eqref{Nalinearm}.
The dynamical part beyond small mass limit can only be obtained through detailed balance condition. As we are going to see, \eqref{Nalinearm} agrees with the result based on detailed balance. 

As a by product, we find the dynamical contribution nicely explains the discrepancy between field theory approaches \cite{Liu:2021uhn,Becattini:2021suc} and kinetic approach \cite{Yi:2021ryh,Hidaka:2017auj} on shear induced spin polarization for massive fermions. In the latter approach, spin polarization is obtained by extrapolating the kinematic contribution in massless theory to massive theory, while in the former approach, both kinematic and dynamical contributions are included implicitly. Turning off the collisional contribution by setting $T_{3p}^{prob}=0$, we find the $O(m)$ term in \eqref{Nalinearm} precisely accounts for the difference between the two approaches in the small mass limit.

\section{Detailed Balance}
\label{sec_staticsolu}
\subsection{collision term}
Now we specify what we mean by detailed balance. For a plasma with steady shear flow, we expect $f_V$ and ${\cal A}^\m$ to reach steady functions \footnote{The condition of steady shear can be relaxed to a slow-varying shear, which does not affect our discussions.}. $f_V$ satisfies the Boltzmann equation. It is known long ago that the presence of a steady shear will drive $f_V$ away from the local equilibrium distribution such that the collision term balance the gradient term on the LHS of the kinetic equation. The situation of ${\cal A}^\m$ is different. ${\cal A}^\m$ satisfies a separate axial kinetic equation, which schematically reads
  \begin{align}
  \label{kineticA}
    P\cdot\pd {\cal A}^\m=C_{\cal A}^\m.
\end{align}
With our counting $f_A\sim O(\pd)$ and $f_V\sim O(\pd^0)$, ${\cal A}^\m\sim O(\pd)$, thus the LHS is counted as $\sim O(\pd^2)$. It follows that the collision term should vanish at $O(\pd)$. This is to be referred as detailed balance condition.

The evolution of spin involves diffusion of the initial spin of the probe fermion as well as polarization induced by the shear flow through collision with medium fermions. Since we are going to use the detailed balance condition of the spin kinetic equation, it is required to evaluate the collision terms up to $\mathcal{O}(\partial)$. By using the complete basis for the Clifford algebra, the Wigner function and the self-energy are decomposed into 
\begin{align}
&S^<=\mathcal{S}+i\mathcal{P}\gamma^5+\mathcal{V}_\mu\gamma^\mu+\mathcal{A}_\mu\gamma^5\gamma^\mu+\frac{1}{2}\mathcal{S}_{\mu\nu}\sigma^{\mu\nu},\nonumber\\
&\Sigma^<=\Sigma_S+i\Sigma_P\gamma^5+\Sigma_{V\mu}\gamma^\mu+\Sigma_{A\mu}\gamma^5\gamma^\mu+\frac{1}{2}\Sigma_{T\mu\nu}\sigma^{\mu\nu}.
\end{align} 
To $\mathcal{O}(\partial)$ order, the collision terms for axial-vector components follow from Kadanoff-Baym equation
\begin{eqnarray}
\label{transportA1}
C_{\cal A}^\m
&=&
-p_\mu \widehat{\Sigma_{A\nu}\mathcal{V}^\nu}
+p_\nu\widehat{\Sigma_{A\mu}\mathcal{V}^\nu}
+m\widehat{\Sigma_S\mathcal{A}_\mu}
+p_\rho\widehat{\Sigma_{V}^\rho \mathcal{A}_\mu}
\nonumber\\
&&
-\frac{1}{2}\epsilon_{\mu\nu\rho\sigma}(\partial^\sigma \widehat{\Sigma_V^{\nu})\mathcal{V}^{\rho}}
+\frac{m}{2}\epsilon_{\rho\sigma\lambda\mu}\widehat{\Sigma_T^{\rho\sigma}\mathcal{V}^\lambda}
+\frac{1}{2}\epsilon_{\rho\sigma\lambda\mu}(\bar{\Sigma}_{V}^\rho+{\Sigma}_{V}^\rho)\widehat{\Sigma_V^{\sigma}\mathcal{V}^{\lambda}},
\end{eqnarray}
where $\widehat{XY}=\bar{X}Y-X\bar{Y}$, $\bar{X}$ and ${X}$ correspond to greater and lesser components respectively. The last term has been ignored in previous studies, since it contains an extra power of $\S_V$ thus is superficially higher order in coupling constant. However, in a steady state, distribution functions in $\widehat{\Sigma_V^{\sigma}\mathcal{V}^{\lambda}}$ should be replaced by $f\rightarrow f_{eq}+\delta f$ with $f_{eq}$ and $\d f$ corresponding to local equilibrium distribution and steady state deviation. The $f_{eq}$ term in $\widehat{\Sigma_V^{\sigma}\mathcal{V}^{\lambda}}$ vanishes by detailed balance, while terms linear in $\delta f$ has additional dependence on coupling constant as \eqref{chi_probe}, which precisely cancels the counterpart in $\widehat{\Sigma_V^{\sigma}\mathcal{V}^{\lambda}}$ \cite{Lin:2022tma}. Besides, since $\delta f$ is first order in gradient, the last term in (\ref{transportA1}) is of $\mathcal{O}(\partial)$ order. Hence the last term in (\ref{transportA1}) contributes at the same order of  coupling constant and gradient as the other terms.

As reasoned earlier, for $m\gg eT$, only Coulomb scattering between probe fermion and medium fermions is relevant to leading logarithmic order.
Fig.\ref{fig1} is the self-energy diagram corresponding to the Coulomb scattering of the massive probe fermion ($P$ and $K$) with the massless medium fermion ($P'$ and $K'$). Writing all the components of the self-energy accordingly, and substituting into (\ref{transportA1}), the collision terms for the Coulomb scattering is 
\begin{eqnarray}
\label{transportn1}
C_{\cal A}^\m&=&-\int_{K,Q,K',P'}\Big\{
M^{A1}_{\mu\nu}\Big(\bar{f}_K {f}_{P'}\bar{f}_{K'}+{f}_K \bar{f}_{P'}{f}_{K'}\Big)N^{\nu}(P)
+M^{A2}_{\mu\nu}\Big({f}_{P'}\bar{f}_{K'}f_{P}+\bar{f}_{P'}{f}_{K'}\bar{f}_{P}\Big)N^\nu(K)\nonumber\\
&&\qquad\qquad\quad\;\,
+M^{A3}_{\mu\nu}\Big(\bar{f}_{K'}f_P\bar{f}_K+{f}_{K'}\bar{f}_P{f}_K\Big)S_u^{\nu\rho}\mathcal{D}_\rho f_{p'}
+M^{A4}_{\mu\nu}\Big({f}_{P'}f_P\bar{f}_K+\bar{f}_{P'}\bar{f}_P{f}_K\Big)S_u^{\nu\rho}\mathcal{D}_\rho f_{k'}\nonumber\\
&&\qquad\qquad\quad\;\,
+M^{A5}_{\mu\nu}\Big(\bar{f}_K{f}_{P'}\bar{f}_{K'}+{f}_K\bar{f}_{P'}{f}_{K'}\Big)\mathcal{D}^\nu {f}_{p}\;\;
+M^{A6}_{\mu\nu}\Big({f}_{P'}\bar{f}_{K'}f_{P}+ \bar{f}_{P'}{f}_{K'}\bar{f}_{P}\Big)\mathcal{D}^\nu{f}_k
\Big\},
\end{eqnarray}
with $\int_{K,Q,K',P'}=2\pi\delta(P^2-m^2)\int\frac{d^4Kd^4Qd^4K'd^4P'}{((2\pi)^4)^4}(2\pi)^8\delta(P-K-Q)\delta(Q+P'-K')\delta(K^2-m^2)\delta(P'^2)\delta(K'^2)$, and effective amplitudes $M^{Ai}_{\mu\nu}$ are given in Appendix (\ref{Mi}). 

The collision term above is aimed to describe the evolution of massive probe spin in a steady shear flow. The first line of (\ref{transportn1}) describes the spin diffusion of the probe fermion, where $N^{\nu}(P)$ and $N^\nu(K)$ are the spin of the massive probe fermion defined in \eqref{Nmu}. The rest of terms in (\ref{transportn1}) are the polarization effect. The second line delineates the polarization contributed from spin of the medium fermion, where $S_u^{\nu\rho}\mathcal{D}_\rho f_{p'}$ and $S_u^{\nu\rho}\mathcal{D}_\rho f_{k'}$ are the spin polarization of the massless medium fermion at local equilibrium \eqref{Amu2}. The third line is the polarization effect induced by inhomogeneous distribution as well as the redistribution of the probe fermion. The collision terms delineate the competition between spin diffusion and polarization. When the evolution of the probe spin reaches a steady state, the diffusion and polarization will balance each other, resulting in a vanishing collision term. 
It is worth noticing that the self-energy correction enters the collision terms through both $S_u^{\nu\rho}\mathcal{D}_\rho f_{p'}$($S_u^{\nu\rho}\mathcal{D}_\rho f_{k'}$) for medium fermions and $\mathcal{D}^\nu f_p$($\mathcal{D}^\nu f_k$) for probe fermion. With these quantities from \eqref{T3med} for medium fermions and from \eqref{T2probe}, \eqref{T3probe} for probe fermions, we are ready to complete the phase space integral and determine $N^{\nu}(P)$ through detailed balance condition.

Before carrying out the phase space integral, it is useful to decompose $C_{\cal A}^\m$. Since $C_{\cal A}^\m$ has the same parity and time-reversal properties as ${\cal A}^\m$, we may as well decompose it as
\begin{eqnarray}
\label{decomposeC}
C_{\cal A\mu}
=
C_{A}L_{\mu,\xi\nu}^P\sigma^{\xi\nu},
\end{eqnarray}
with $L^{\mu,\xi\lambda}_P$ defined in \eqref{projector}. $C_{A}$ is a scalar function of momentum, it is obtained from the collision term \eqref{transportn1} with straightforward but tedious calculation  
\begin{eqnarray}
\label{CA}
C_{A}&=&16N_fe^4(2\pi)^3\int_{K,Q,K',P'}\frac{1}{(Q^2)^2}\nonumber\\
&\times&\bigg\{\frac{J_1}{2P\cdot uP_\perp^2}
\Big(N_P-1+\frac{2T_{3p}^{\text{prob}}}{p}\Big)
+\frac{J_2}{2K\cdot uP_\perp^2}\Big(N_K-1+\frac{2T_{3k}^{\text{prob}}}{k}\Big)\nonumber\\
&&+\frac{m^2}{2K\cdot uP_\perp^2}\Big(J_3 N_K-J_4(1-\frac{2T_{3k}^{\text{prob}}}{k})-J_5\frac{T_{2k}^{\text{prob}}}{k}\Big)+\frac{m^2J_6}{2P\cdot uP_\perp^2}\Big(1-\frac{2T_{3p}^{\text{prob}}}{p}\Big)\nonumber\\
&&+\frac{m^2J_7}{2K'\cdot u P_\perp^2}\Big(1-\frac{2T^{\text{med}}_{3k'}}{k'}\Big)-\frac{m^2J_{8}}{2P'\cdot u P_\perp^2}\Big(1-\frac{2T^{\text{med}}_{3p'}}{p'}\Big)\bigg\}\bar{f}_k \bar{f}_{k'}{f}_{p'}f_p,
\end{eqnarray}
where the $J_i$ are functions of momentum $P,K,P',K'$ defined in \eqref{Ji}, $T_2^{\text{prob}}$, $T_3^{\text{prob}}$ and $T_3^{\text{med}}$ comes from the self-energy correction with their explicit expressions given in \eqref{T2probe}, \eqref{T3probe} and \eqref{T3med}. The task of determining the local equilibrium $\mathcal{A}^\mu$ from the detailed balance is now turned into finding the coefficient $N_P$ that eliminate $C_{A}$.

In the massless limit $C_{A}=0$ has an exact solution, with terms proportional to $m^2$ vanishing in \eqref{CA}, one can directly read from \eqref{CA} that 
\begin{eqnarray}
\label{masslesssolution}
N_P=1-2T_{3p}^{\text{prob}}/p. 
\end{eqnarray}
From (\ref{NPdecompose}), such solution of massless fermion indicates $N_a=0$, in other word $P_\mu f_A=0$ in the massless case. This agrees with our analysis in the Sec.\ref{sec_frame} and \cite{Hidaka:2017auj, Fang:2022ttm}. For massive fermion, there is no such simple solution, however, the vanishing of collision term gives a constraint on $N_P$ and $N_K$, which leads to a complicated integral equation. Fortunately, it can be simplified to a differential equation in the leading logarithmic order approximation: With the probe fermion and medium fermion being hard fermions, and the momentum transfer being soft $eT\ll q_0,q\ll T$, $N_K$ can be expanded around $N_P$ as
\begin{eqnarray}
N_K=N_{P-Q}= N_P-Q^\nu\partial_{P^\nu} N_P+\frac{1}{2}Q^\rho\partial_{P^\rho}Q^\nu\partial_{P^\nu}N_P+\mathcal{O}(Q^3).
\end{eqnarray}
Since $N_P$ is only a function of $|\vec{p}|$, the expansion above can be further simplified to be 
\begin{eqnarray}
\label{NKexpand}
N_K=N_P-q \cos\theta_q {N'_P}+\frac{1}{2p}(\sin^2\theta_q {N'_P}+\cos^2\theta_q {N''_P})q^2+\mathcal{O}(q^3),
\end{eqnarray}
with $\cos\theta_q\equiv\hat{q}\cdot\hat{p}$ and $N'_P=\partial_p N_P$. The calculation details of momentum integral are presented in Appendix.\ref{details}. After momentum integral, the detailed balance condition is converted into a differential equation of $N_P$
\begin{eqnarray}
\label{differential}
C_{A}=c_\text{diff}^{(0)} N_P+c_\text{diff}^{(1)} {N'_P}+c_\text{diff}^{(2)} {N''_P}+ c_\text{pol}=0,
\end{eqnarray}
where $c_\text{diff}^{(0)} N_P+c_\text{diff}^{(1)} {N'_P}+c_\text{diff}^{(2)} {N''_P}$ is the diffusion of probe spin, leading to decrease of the probe spin with $c_\text{diff}^{(0)}<0$. $c_\text{pol}$ is the polarization effect, which contains contribution from different parts $c_\text{pol}=c_\text{pol}^{\text{med}\partial}+ c_\text{pol}^{\text{med}\Sigma}+ c_\text{pol}^{\text{prob}\partial}+ c_\text{pol}^{\text{prob}\Sigma}$. Definitions and explicit expressions of all parts are given in Appendix.\ref{details}. $c_\text{pol}^{\text{med}\partial}$ is polarization effect contributed from the magnetization current of the medium fermion, $c_\text{pol}^{\text{med}\Sigma}$ from the displacement current of the medium fermion, $c_\text{pol}^{\text{prob}\partial}$ from the inhomogeneity of the probe distribution, and $c_\text{pol}^{\text{prob}\Sigma}$ from the redistribution of the probe fermion in the shear flow. 

The competition between polarization and diffusion rate can be characterized through the ratio $-c_\text{pol}/c_\text{diff}^{(0)}$, which depends on temperature, momentum, and mass of the probe. The ratio between polarization and diffusion can be simplified in certain limit. In the relativistic limit $p\gg m$, the ratio between polarization and diffusion effect approaches
\begin{eqnarray}
\label{poltodiff1}
-\frac{c_\text{pol}}{c_\text{diff}^{(0)}}\bigg|_{p\gg m}=\frac{3}{2},
\end{eqnarray}
wherein the polarization effect from the medium fermion does not contribute in such limit $c_\text{pol}^{\text{med}\partial}/c_\text{diff}^{(0)}\rightarrow 0$, $c_\text{pol}^{\text{med}\Sigma}/c_\text{diff}^{(0)}\rightarrow 0$, while that from the inhomogeneity of probe fermion $-c_\text{pol}^{\text{prob}\partial}/c_\text{diff}^{(0)}\rightarrow 1$, $-c_\text{pol}^{\text{prob}\Sigma}/c_\text{diff}^{(0)}\rightarrow 1/2$. In the non-relativistic limit $m\gg p$, the ratio between polarization and diffusion effect is 
\begin{eqnarray}
\label{poltodiff2}
-\frac{c_\text{pol}}{c_\text{diff}^{(0)}}\bigg|_{m\gg p}&=&\frac{1}{2}+\Big(\frac{\pi ^2 (C_b-C_f)}{16}-C_fN_f\Big(\frac{9  \zeta (3)}{4 \pi ^2}+ \ln 2\Big)\Big)\nonumber\\
&&+\frac{T (7 m-225  C_fN_f T \zeta (3))}{50 p^2}+\tanh\left(\frac{m}{2 T}\right) \left(\frac{173}{600}+\frac{21 C_fN_f T \zeta (3)}{5m}\right)-\frac{128 T}{35 m}+\frac{141C_fN_f  \zeta (3) T^2}{5m^2}.
\end{eqnarray}
The magnetization current of the medium fermion spin $-c_\text{pol}^{\text{med}\partial}/c_\text{diff}^{(0)}\rightarrow 1/4$ and inhomogeneity of probe fermion $-c_\text{pol}^{\text{prob}\partial}/c_\text{diff}^{(0)}\rightarrow 1/4$ make the first $1/2$ in the first line. The second and third terms in the first line come from the displacement current of the medium fermion $c_\text{pol}^{\text{med}\Sigma}$, which also approaches a constant in such limit. The redistribution of the probe fermion $c_\text{pol}^{\text{prob}\Sigma}$ contributes to the second line, such effect is linear in $m$ in the large mass limit. Coefficients in \eqref{differential} under both limits are presented in the Appendix.~\ref{details}

\subsection{boundary condition and solution}
In order to solve the differential equation \eqref{differential}, boundary conditions are required. We first work out the boundary condition. With the coefficients in the large momentum limit $p\gg m$ \eqref{coe_largep}, the corresponding differential equation can be solved analytically, keeping leading and subleading terms, the solution of the differential equation becomes 
\begin{eqnarray}
\label{largepsolution}
N_P|_{p\gg m}&=&\frac{3}{2}+\frac{21T}{2p}. 
\end{eqnarray}
The homogeneous solution is discarded since without the inhomogeneous term sourced by shear, detailed balance for axial component would favor a vanishing axial component for the Wigner function. The solution above agrees with our analysis from \eqref{CA} upon using $T_{3p}^{\text{prob},m=0}$ \eqref{T3massless} in the large momentum limit. 

On the other hand, with the coefficients in the small momentum limit $p\ll m$ \eqref{coe_smallp}, the analytical solution of the differential equation reads 
\begin{eqnarray}
\label{nonrelasolution}
N_P|_{p\ll m}&=&\frac{T \left(7 m-225 T \zeta (3) C_f N_f\right)}{50 p^2}+\frac{1}{2}+\Big(\frac{\pi ^2 (C_b-C_f)}{16}-C_fN_f\Big(\frac{9  \zeta (3)}{4 \pi ^2}+ \ln 2\Big)\Big) \nonumber\\
&&+\tanh
   \left(\frac{m}{2 T}\right) \left(\frac{27 T  C_f N_f\zeta (3)}{10 m}+\frac{67}{200}\right)-\frac{38789 T}{10500 m}+\frac{3087 C_f N_f  \zeta (3) T^2}{105 m^2}, 
\end{eqnarray}
where we keep the leading and subleading order. We see the enhancement of polarization in the infrared (IR) as $N_p\sim p^{-2}$. This can be traced back to the corresponding term in \eqref{poltodiff2}. As remarked before, this comes from collisional effect through redistribution of the probe fermion. An enhancement of the same origin has been found in the massless counterpart \cite{Lin:2024zik}.
\begin{figure}[H]\centering
\includegraphics[height=0.3\textwidth]{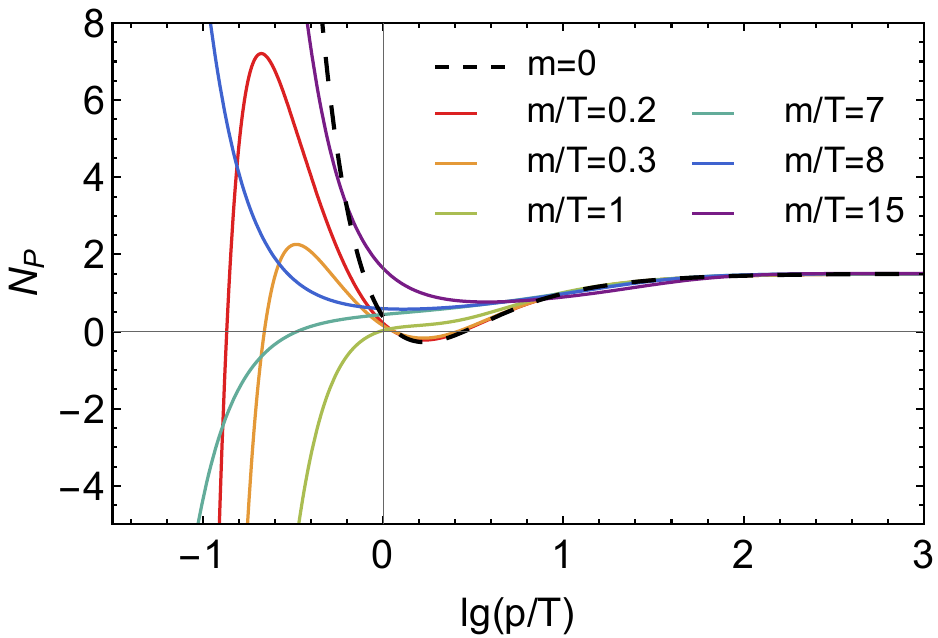}
\caption{$N_P$ defined in \eqref{defineNP} as a function of $\text{log}_{10}(p/T)$ for different masses.}
\label{Np}
\end{figure}
Between both limits, the differential equation can only be solved numerically. We take analytical solutions above as the boundary condition for numerically solving the differential equation \eqref{differential}. The differential equation for two flavor medium fermion $N_f=2$ is solved using shooting method. Numerical solution $N_P$ \eqref{defineNP} as a function of $p/T$ for different masses is presented in Fig.\ref{Np}. The black dashed line is the exact solution \eqref{masslesssolution} in the massless limit, the colored lines are solutions with different masses. In the large momentum limit $p\gg m$ and also $p\gg T$, the solutions all approaches $3/2$ as fixed by our boundary condition \eqref{largepsolution}. For moderate $p$, the curves deviate from each other. For $p\lesssim T$, the solution becomes unreliable. An input for the redistribution of probe fermions beyond the approximate solution \eqref{chi_probe} is needed.

\begin{figure}[H]\centering
\includegraphics[height=0.3\textwidth]{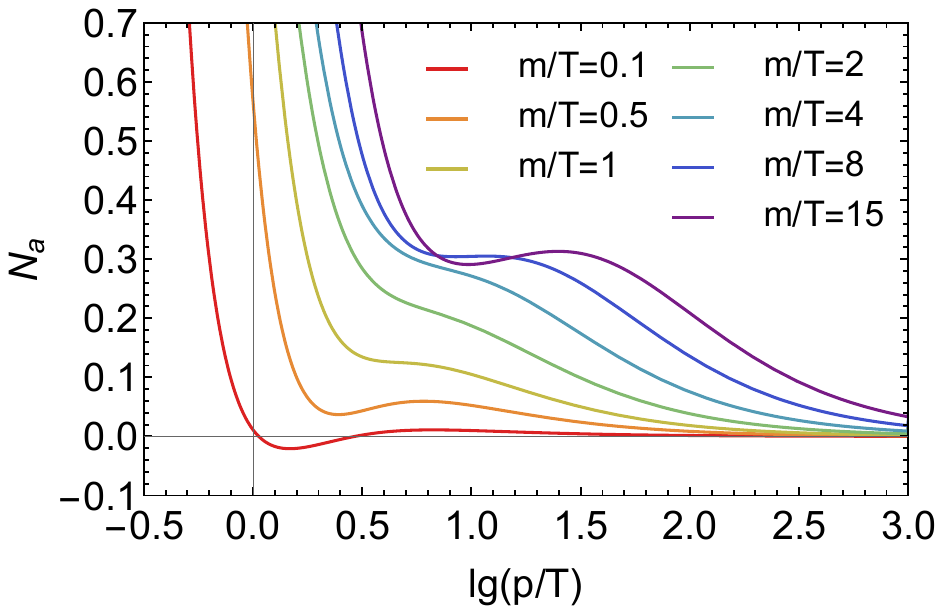}
\caption{The dynamical part $N_a$ as a function of $\text{log}_{10}(p/T)$ for different masses. }
\label{Na}
\end{figure}
Recall that from the analysis \eqref{NPdecompose}, $N_P$ is the sum of dynamical contribution $N_a$, kinematic contribution $N_\partial$ and collisional contribution $N_\Sigma$. We have calculated $N_\partial$ and $N_\Sigma$ in the Sec.\ref{sec_selfenergy}, subtracting these two terms from $N_P$ will give the dynamical part $N_a$. We plot the dynamical part $N_a$ as a function of momentum in Fig.\ref{Na}. It shows that dynamical part $a_\mu f_A$ gets strongly suppressed at large momentum, and $a_\mu f_A\rightarrow 0$ when $p/T\rightarrow\infty$. In the massless limit $N_a$ always vanishes, this is in consistency with  analysis \eqref{masslesssolution}. For relatively large mass, $a_\mu f_A$ increases almost linearly with $m/T$. 
\begin{figure}[H]\centering
\includegraphics[height=0.3\textwidth]{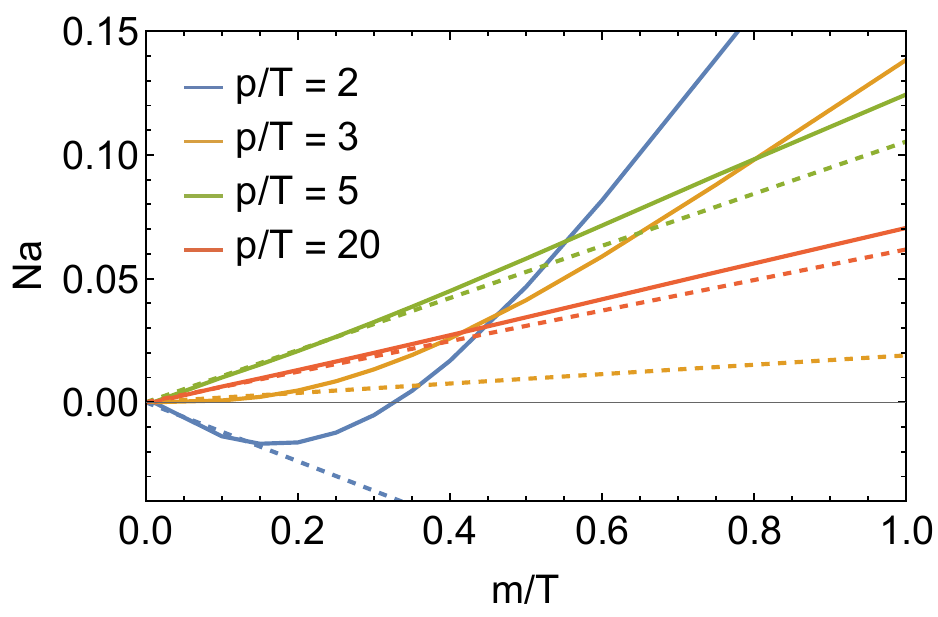}
\includegraphics[height=0.3\textwidth]{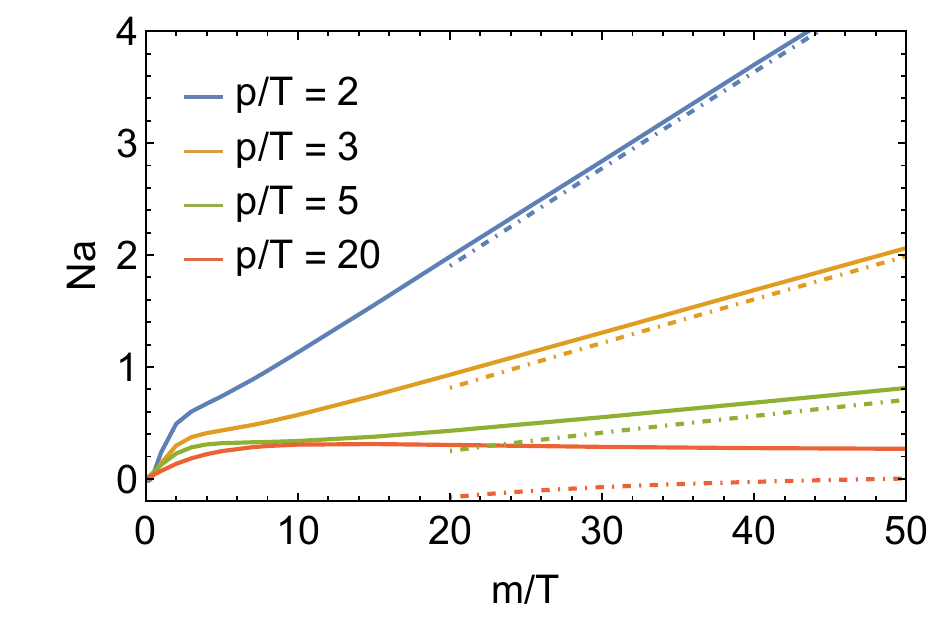}
\caption{The dynamical part $N_a$ as a function of $m/T$ for small masses (left) and large masses (right). Solid lines are from numerical solution to \eqref{differential}. Dashed lines and dot-dashed lines correspond to \eqref{Nalinearm} and \eqref{nonrelasolution} respectively.}
\label{Naofm}
\end{figure}
The dependence of the dynamical part $N_a$ on mass is presented in Fig.\ref{Naofm}. In the left panel, we present $N_a$ at small mass for various momentum. The solid lines are the numerical solution, while the dashed lines give a comparison with $N_a$ obtained through frame independence \eqref{Nalinearm}, namely 
\begin{eqnarray}
N_a|_{p\gg m}=\frac{m}{p}\Big(1-\frac{2T_{3p}(m=0)}{p}\Big).
\end{eqnarray}
Clearly, in the limit $p\gg m$, \eqref{Nalinearm} is a good approximation to $N_a$. In the right panel, numerical solutions to $N_a$ in a wide range of mass are presented as the solid lines. The dot-dashed lines represent the analytical approximation of $N_a$ in the case $m\gg p$. This approximation is obtained through substracting $N_\partial+N_\Sigma$ from \eqref{nonrelasolution} and taking the corresponding limit, giving
\begin{eqnarray}
N_a|_{m\gg p}=\frac{17 m T}{50 p^2}+C_f N_f \left(\frac{9}{4} \zeta (3) \left(\frac{2 T^2}{p^2}-\frac{1}{\pi ^2}\right)-\ln2\right)+\frac{241}{1200}.
\end{eqnarray}
As is shown in the figure, both lines agree when $m\gg p$.

Now we discuss implication for spin polarization in heavy ion collisions. 
Despite the complicated dependence of $N_P$ on $m$ and $p$, we see from Fig.~\ref{Np} that the complete polarization is in general larger than $N_P=1$ obtained ignoring collisions \cite{Liu:2021uhn,Becattini:2021suc,Hidaka:2017auj}. If we naively apply the QED results to QCD, taking $p$ to be a few GeV, $T$ to be the freezeout temperature and fermion mass to be $\sim100 \text{MeV}$ for the strange quark, we seem to be in the limit $p\gg T$ and $p\gg m$, in which $N_P\to 3/2$. It indicates that collisional effect ignored in previous phenomenological studies actually enhances the spin-shear coupling by $50\%$! This is in contrast to the suppression found by us in \cite{Lin:2022tma}. The difference follows from different assumptions on distribution of the probe fermion: in the present work, the probe fermion is in steady state, while in \cite{Lin:2022tma}, the probe fermion is assumed to be in local equilibrium.
The enhancement seems to be phenomenologically favored as existing studies suggest that the collisionless spin-shear coupling may not be sufficient to account for the local spin polarization \cite{Becattini:2021iol,Yi:2021ryh}.
The enhancement seems to be phenomenologically favored as existing studies suggest that the collisionless spin-shear coupling may not be sufficient to account for the local spin polarization \cite{Becattini:2021iol,Yi:2021ryh}.

\section{Conclusion and Outlook}
\label{s5}
In the QKT framework, the spin polarization ${\cal A}^\m$ is composed of the dynamical term $a^\mu f_A$ and a non-dynamical part $S_{m,n}^{\mu\nu}\mathcal{D}_\nu f$. The non-dynamical part including collisional contribution in a shear flow has been obtained in \cite{Lin:2022tma,Lin:2024zik}. The dynamical part is unique to massive theory and involves solving the axial kinetic equation governing the evolution of spin. Power counting indicates that the axial-vector component of Wigner function satisfies a detailed balance condition, in which the corresponding component of collision term vanishes. This allows us to fix ${\cal A}^\m$ as a whole, which can then be split into three parts with existing knowledge about non-dynamical part.

We perform analysis with the principle of frame independence of ${\cal A}^\m$ in the small mass regime, which enables us to constrain the form of dynamical part of ${\cal A}^\m$. The resulting dynamical part in the fluid frame explains the discrepancy in spin polarization between field theory approach and kinetic theory approach in the corresponding regime in the collisionless limit. We also perform explicit calculations for a massive probe fermion undergoing Coulomb scattering in a massless QED plasma with a steady shear flow, aiming at understanding strange quark polarization in QGP. We find the corresponding solution to $\mathcal{A}_\mu$ and $a^\mu f_A$. In the phenomenologically interesting region, we find the collisional effect enhances the spin-shear coupling by $50\%$, which may help better understand the measurement of local spin polarization of $\L$ hyperon in heavy ion collisions.

For phenomenological application, generalizations of the present work from QED to QCD case is necessary. In a QCD plasma, additional gluon loop presents in the collision terms, hence the redistribution of both quarks and gluons would contribute. We leave this for future studies.



\acknowledgments
We are grateful to Shi Pu and Di-Lun Yang for fruitful discussions. The work is supported by NSFC grant Nos. 12005112 (Zy.W.) and Nos 12075328, 11735007 (S.L.).

\begin{appendix}
\section{Calculation details}
\label{details}
Effective amplitudes $M^{Ai}_{\mu\nu}$ in (\ref{transportn1}) are given by 
\begin{eqnarray}
\label{Mi}
M^{A1}_{\mu\nu}&=&\frac{8N_fe^4(2\pi)^3}{(Q^2)^2}2\Big(-m^2P'\cdot K'+P\cdot P' K\cdot K'+K\cdot P' P\cdot K'\Big)g_{\mu\nu},\nonumber\\
M^{A2}_{\mu\nu}&=&\frac{8N_fe^4(2\pi)^3}{(Q^2)^2}\Big\{2\Big(P'\cdot K' P\cdot K-P\cdot P' K'\cdot K-P\cdot K' P'\cdot K\Big)g_{\mu\nu}\nonumber\\
&&\qquad\qquad\quad\;-2P_\mu(P\cdot P' K'_\nu +P\cdot K' P'_\nu )
+2K_\mu (P\cdot P' K'_\nu +P\cdot K' P'_\nu-P'\cdot K' P_\nu )\nonumber\\
&&\qquad\qquad\quad\;
+2P'_\mu (K'\cdot K P_\nu -(P\cdot K-m^2) K'_\nu )
+2K'_\mu (P'\cdot K P_\nu -(P\cdot K-m^2) P'_\nu )\Big\},\nonumber\\
M^{A3}_{\mu\nu}&=&\frac{8N_fe^4(2\pi)^3}{(Q^2)^2}2\Big(m^2Q\cdot K' g_{\mu\nu}-m^2K'_\mu Q_\nu +K\cdot  K' P_\mu  P_\nu -P\cdot  K' P_\mu K_\nu\Big),\nonumber\\
M^{A4}_{\mu\nu}&=&\frac{8N_fe^4(2\pi)^3}{(Q^2)^2}2\Big(m^2Q\cdot P' g_{\mu\nu}-m^2P'_\mu Q_\nu+ K\cdot P' P_\mu P_\nu-P\cdot  P' P_\mu K_\nu\Big),\nonumber\\
M^{A5}_{\mu\nu}&=&-\frac{8N_fe^4(2\pi)^3}{(Q^2)^2}\epsilon_{\mu\nu\rho\sigma}P^{\rho}\Big(K\cdot K'P'^{\sigma} +K\cdot P' K'^{\sigma}\Big),\nonumber\\
M^{A6}_{\mu\nu}&=&\frac{8N_fe^4(2\pi)^3}{(Q^2)^2}\Big\{\epsilon_{\mu\nu\rho\sigma}K^{\rho} \Big(P\cdot K' P'^{\sigma}+P\cdot P' K'^{\sigma} -P'\cdot K' P^{\sigma}\Big)+\epsilon_{\nu\rho\sigma\lambda}\Big(P'_\mu K'^{\sigma}+K'_\mu P'^{\sigma}\Big)P^{\lambda}K^{\rho}\Big\}.
\end{eqnarray}
Additional factor $2N_f$ comes from scattering with $N_f$ fermions and anti-fermions. 
The $J_i$ in \eqref{CA} are functions of momentum $P,K,P',K'$ defined below
\begin{eqnarray}
\label{Ji}
J_1&=&(P\cdot P'K\cdot K'+K\cdot P'P\cdot K'-m^2P'\cdot K')2P_\perp^4,\nonumber\\
J_2&=&-(P\cdot P' K'\cdot K+P\cdot K' P'\cdot K-P'\cdot K' P\cdot K)\big(3(P_\perp\cdot K_\perp)^2-K_\perp^2P_\perp^2\big)\nonumber\\
&&-P'\cdot K' P_\perp\cdot K_\perp\big((P_\perp\cdot K_\perp)^2-K_\perp^2P_\perp^2\big)\nonumber\\
&&+\big\{K\cdot K'(P_\perp\cdot K_\perp)((P'_\perp\cdot P_\perp)(P_\perp\cdot K_\perp)-(P'_\perp\cdot K_\perp)P_\perp^2)\big\}+\{P'\leftrightarrow K'\}\nonumber\\
&&+\big\{P\cdot K'(P_\perp\cdot K_\perp)((P'_\perp\cdot K_\perp)(P_\perp\cdot K_\perp)-(P'_\perp\cdot P_\perp)K_\perp^2)\big\}+\{P'\leftrightarrow K'\}\nonumber\\
&&-\big\{P\cdot K \big[(\vec{k}'\times\vec{p}\cdot\vec q)^2+(P'_\perp\cdot K'_\perp)(P_\perp\cdot K_\perp)^2-(P'_\perp\cdot K_\perp)(K'_\perp\cdot P_\perp)(P_\perp\cdot K_\perp)\big]\big\}-\{P'\leftrightarrow K'\},\nonumber\\
J_3&=&\big\{(\vec{k}'\times\vec{p}\cdot\vec q)^2+(P'_\perp\cdot K'_\perp)(P_\perp\cdot K_\perp)^2-(P'_\perp\cdot K_\perp)(K'_\perp\cdot P_\perp)(P_\perp\cdot K_\perp)\big\}+\{P'\leftrightarrow K'\},\nonumber\\
J_4&=&\big\{(P_\perp\cdot K_\perp)[(P'_\perp\cdot K_\perp)P_\perp^2-(P'_\perp\cdot K_\perp)(K'_\perp\cdot P_\perp)]\big\}+\{P'\leftrightarrow K'\},\nonumber\\
&&-\big\{P'\cdot K'(P_\perp\cdot K_\perp)P_\perp^2+P\cdot K'(P'_\perp\cdot K_\perp)P_\perp^2-3P\cdot K'(P'_\perp\cdot P_\perp)(P_\perp\cdot K_\perp)\big\}-\{P'\leftrightarrow K'\},\nonumber\\
J_5&=&4(\vec{k}'\times\vec{p}\cdot\vec q)^2(P_\perp\cdot K_\perp)/K_\perp^2,\nonumber\\
&&-\big\{[2P\cdot K'(P_\perp\cdot P'_\perp K_\perp^2-P_\perp\cdot K_\perp P'_\perp\cdot K_\perp)+P'\cdot K'((P_\perp\cdot K_\perp)^2-P_\perp^2K_\perp^2)](P_\perp\cdot K_\perp)/K_\perp^2\big\}-\{P'\leftrightarrow K'\},\nonumber\\
J_6&=&2P_\perp^2[K\cdot K'(P'_\perp\cdot P_\perp)+K\cdot P'(K'_\perp\cdot P_\perp)-P'\cdot K'P_\perp^2]\nonumber\\
J_7&=&P'\cdot K'(3(K'_\perp\cdot P_\perp)^2-{K'_\perp}^2P_\perp^2)+(\vec{k}'\times\vec{p}\cdot\vec q)^2+{P'_\perp}^2(K'_\perp\cdot P_\perp)^2-(P'_\perp\cdot K'_\perp)(K'_\perp\cdot P_\perp)(P'_\perp\cdot P_\perp),\nonumber\\
J_{8}&=&P'\cdot K'(3(P'_\perp\cdot P_\perp)^2-{P'_\perp}^2P_\perp^2)+(\vec{k}'\times\vec{p}\cdot\vec q)^2+{K'_\perp}^2(P'_\perp\cdot P_\perp)^2-(P'_\perp\cdot K'_\perp)(P'_\perp\cdot P_\perp)(K'_\perp\cdot P_\perp),
\end{eqnarray}
where $\{P'\leftrightarrow K'\}$ means exchanging the two momentum in the bracket $\{\cdots\}$ ahead. 

Assuming that the medium fermion and probe fermion are hard fermions with momentum comparable with temperature $p, k, p', k'\sim T$, while the momentum transfer is soft $q_0, q\ll T$, the phase space integral can be simplified using the small momentum transfer as well as momentum conservation and on-shell condition
\begin{eqnarray}
&&(2\pi)^3\int\frac{d^4Kd^4Qd^4P'd^4K'}{((2\pi)^4)^4}(2\pi)^8\delta(P-K-Q)\delta(Q+P'-K')\delta(K^2-m^2)\delta(P'^2)\delta(K'^2)\nonumber\\
&=&\frac{1}{(2\pi)^5}\int dq_0 d^3qd^3k'\frac{1}{2p'_02k'_02k_0}\delta(p_0-k_0-q_0)\delta(p'_0-k'_0+q_0).
\end{eqnarray}
The momentum integral is left with integral over $Q$ and $\vec k'$. It is useful to decompose momentum $\vec q$ and $\vec k'$ into
\begin{eqnarray}
\label{momentum_decomposition}
\vec{k}'&=& k'\cos\theta_k\hat{p}+k'\sin\theta_k\cos\varphi_k\hat{x}+k'\sin\theta_k\sin\varphi_k\hat{y},
\nonumber\\
\vec{q}&=& q\cos\theta_q\hat{p}+q\sin\theta_q\cos\varphi_q\hat{x}+q\sin\theta_q\sin\varphi_q\hat{y},
\end{eqnarray}
where we have denoted $\hat{p}$ as $\hat{z}$ for now. We introduce $\Omega$ as the angle between $\vec k'$ and $\vec q$, namely $\cos\Omega=\cos\theta_k\cos\theta_q-\sin\theta_k\sin\theta_q\cos\Delta\varphi$, with $\Delta\varphi=\varphi_q-\varphi_k$.  The measure can be parametrized as 
\begin{eqnarray}
\int d^3qd^3k'=\int q^2 dq d\cos\theta_q d\varphi_{q} k'^2 dk' d\cos\theta_k d\Delta\varphi.
\end{eqnarray}
Considering that loop fermions are light fermions, they can be treated as massless. 
Using $\vec p'=\vec k'-\vec q$ and $\vec k=\vec p-\vec q$, we can use the on-shell condition to cast the $\delta$-function into 
\begin{eqnarray}
\delta(p_0-k_0-q_0)&\simeq&\delta(q \frac{p}{p_0}\cos\theta_q-q^2\frac{(p^2\sin^2\theta_q+m^2)}{2p_0^3}-q_0),\nonumber\\
\delta(p'_0-k'_0+q_0)&\simeq&\delta(q\cos\Omega-q^2\frac{\sin^2\Omega}{2k'}-q_0),
\end{eqnarray}
with $p_0=(p^2+m^2)^{1/2}$. The angular integral over $\varphi_{q}$ and $\varphi_{k}$ can be performed to obtain, 
\begin{eqnarray}
\int d\varphi_{q}d\varphi_{k}\delta(p'_0-k'_0+q_0)&\simeq&4\pi\frac{1}{q(1+\frac{q_0}{k'})}\frac{1}{[\sin^2\theta_q\sin^2\theta_k-(\cos\Omega-\cos\theta_q\cos\theta_k)^2]^{1/2}}.
\end{eqnarray}
Note that the above $\delta-$function constrain the unique solution of $\cos\Delta\varphi$, yet $\sin\Delta\varphi$ can take both solutions $\pm(1-\cos^2\Delta\varphi)^{1/2}$. Thus integrals containing odd number of $\sin\Delta\varphi$ will be vanishing under the angular integral. The square root constrains the domain of $\cos\theta_k$ as $\cos(\theta_q-\Omega)<\cos\theta_k<\cos(\theta_q+\Omega)$. The other $\delta$-function gives 
\begin{eqnarray}
\int d\cos\theta_q\delta(p_0-k_0-q_0)\simeq\frac{1}{\frac{pq }{p_0}(1+\frac{q_0}{p_0})}.
\end{eqnarray}
From the $\delta$-function, one can solve 
\begin{eqnarray}
&&\cos\Omega\simeq\frac{q_0}{q}+\frac{q}{2k'}\Big(1-\frac{q_0^2}{q^2}\Big)+\mathcal{O}(q^2),\qquad\qquad\quad
\sin\Omega\simeq\Big(1-\frac{q_0^2}{q^2}\Big)^{1/2}\Big(1-\frac{q_0}{2{k'}}\Big),\nonumber\\
&&\cos\theta_q\simeq\frac{p_0q_0}{pq}+\frac{q}{2p}\Big(1-\frac{q_0^2}{q^2}\Big)
+\mathcal{O}(q^2),\qquad\qquad 
\sin\theta_q\simeq\Big(1-\frac{p_0^2}{p^2}\frac{q_0^2}{q^2}\Big)^{1/2}\Big(1-\frac{q^2-q_0^2}{p^2-p_0^2q_0^2}\frac{q_0p_0}{2}\Big),
\end{eqnarray}
in obtaining the leading-log order result, it is enough to keep the above solution to the first order of $q$. Note that $-1\leq \cos\Omega,\cos\theta_q\leq 1$ also set a limit to $x=q_0/q$ that $-\frac{p}{p_0}\leq\frac{q_0}{q}\leq\frac{p}{p_0}$. So that $\int dx dk'$ has the integration domain $\int dx dk'\rightarrow   \int_0^\infty dk'\int_{-p/p_0}^{+p/p_0} dx$. With the above approximation of small momentum transfer, the collision term at leading logarithmic order can be explicitly calculated. The basic process is to collect all terms of integrad to $Q^{-2}$, after combining the measure, and integrating $q$ ranges from $eT\ll q\ll T$, the log then arises from $\int_{eT}^{T}dq/q=\ln(1/e)$. 
 
 The various terms in the differential equation $c_\text{diff}^{(0)} N_P+c_\text{diff}^{(1)} {N'_P}+c_\text{diff}^{(2)} {N''_P}+ c_\text{pol}=0$ are defined as follows. $c_\text{diff}^{(0)} N_P+c_\text{diff}^{(1)} {N'_P}+c_\text{diff}^{(2)} {N''_P}$ is the diffusion of probe spin, which is defined as the corresponding parts in \eqref{CA} by
\begin{eqnarray}
c_\text{diff}^{(0)} N_P+c_\text{diff}^{(1)} {N'_P}+c_\text{diff}^{(2)} {N''_P}=16N_fe^4(2\pi)^3\int_{K,Q,K',P'}\frac{1}{(Q^2)^2}\bigg\{\frac{J_1}{2P\cdot uP_\perp^2}
N_P
+\frac{(J_2+m^2J_3)}{2K\cdot uP_\perp^2}N_K\bigg\}\bar{f}_k \bar{f}_{k'}{f}_{p'}f_p,
\end{eqnarray}
due to the small momentum transfer, $N_K$ is expanded around $N_P$ by \eqref{NKexpand}. After performing the phase space integral, the coefficients $c_\text{diff}^{(i)}$ are given as the following, 
\begin{eqnarray}
c_\text{diff}^{(0)}&=&c_A\frac{\left(p^2-2 p_0^2\right)  p^3 p_0^3F+\left(-2
   p^6+6 p_0^2 p^4-11 p_0^4 p^2+3 p_0^6\right)p  T-m^2 p_0 \eta _p \left(F p_0 \left(p^2-2 p_0^2\right) p^2+\left(2 p^4-5 p_0^2 p^2+3 p_0^4\right) T\right)}{p^3 p_0^4},\nonumber\\
c_\text{diff}^{(1)}&=&c_A\frac{-F p^3 p_0^3+p T \left(m^2 p_0^2+2 p^4+2 p_0^4\right)+m^2 p_0 \eta _p \left(F p_0 p^2+\left(p^2-3 p_0^2\right) T\right)}{p^2 p_0^2},\nonumber\\
c_\text{diff}^{(2)}&=&c_A\frac{p_0 T \left(p p_0-m^2 \eta _p\right)}{p},
\end{eqnarray}
where $c_A=\frac{T^2 N_f}{24 \pi }$ is an overall constant. $c_\text{pol}$ is the polarization effect, which contains contribution from different parts $c_\text{pol}=c_\text{pol}^{\text{med}\partial}+ c_\text{pol}^{\text{med}\Sigma}+ c_\text{pol}^{\text{prob}\partial}+ c_\text{pol}^{\text{prob}\Sigma}$. Where $c_\text{pol}^{\text{med}\partial}$ is the polarization effect contributed from the magnetization current of the medium fermion, the definition and expression are 
\begin{eqnarray}
c_\text{pol}^{\text{med}\partial}&=&16N_fe^4(2\pi)^3\int_{K,Q,K',P'}\frac{1}{(Q^2)^2}\bigg\{\frac{m^2J_7}{2K'\cdot u P_\perp^2}-\frac{m^2J_{8}}{2P'\cdot u P_\perp^2}\bigg\}\bar{f}_k \bar{f}_{k'}{f}_{p'}f_p\nonumber\\
&=&c_A\frac{m^2 \left(F p_0^2 \left(m^2+2 p_0^2\right) \eta _p-3 F p_0^3 p+2 p^3 T\right)}{2 p^3 p_0^2}.
\end{eqnarray}
$c_\text{pol}^{\text{med}\Sigma}$ from the displacement current of the medium fermion,
\begin{eqnarray}
c_\text{pol}^{\text{med}\Sigma}&=&16N_fe^4(2\pi)^3\int_{K,Q,K',P'}\frac{1}{(Q^2)^2}\bigg\{-\frac{m^2J_7}{2K'\cdot u P_\perp^2}\frac{2T^{\text{med}}_{3k'}}{k'}+\frac{m^2J_{8}}{2P'\cdot u P_\perp^2}\frac{2T^{\text{med}}_{3p'}}{p'}\bigg\}\bar{f}_k \bar{f}_{k'}{f}_{p'}f_p\nonumber\\
&=&-\frac{9 c_AC_fN_f m^2 \zeta (3) \left[(F-2) p_0^2 (3 p_0^2-p^2) \eta _p-3 (F-2) p_0^3 p+2 p^3 T\right]}{2 \pi ^2 p^3 p_0^2}\nonumber\\
&&+\frac{2c_AC_fN_f m^2 \ln2 \left[F p_0^2 \left(p^2-3 p_0^2\right) \eta _p+3 F p_0^3 p-2 p^3 T\right]}{p^3 p_0^2}\nonumber\\
&&+\frac{\pi ^2 c_A(C_b-C_f) m^2 \left[p_0^2 \left(3 p_0^2-p^2\right) (2 F+1+2\ln2) \eta _p-3 p_0^3 p (2 F+1+2\ln2)+4 p^3 T\right]}{16 p^3 p_0^2}.
\end{eqnarray}
$c_\text{pol}^{\text{prob}\partial}$ is the contribution from the inhomogeneity of the probe distribution, 
\begin{eqnarray}
c_\text{pol}^{\text{prob}\partial}&=&16N_fe^4(2\pi)^3\int_{K,Q,K',P'}\frac{1}{(Q^2)^2}\bigg\{\frac{m^2J_6-J_1}{2P\cdot uP_\perp^2}
-\frac{m^2J_4+J_2}{2K\cdot uP_\perp^2}
\bigg\}\bar{f}_k \bar{f}_{k'}{f}_{p'}f_p\nonumber\\
&=&c_A\frac{\left(9 m^4+7 m^2 p^2+2 p^4\right)p_0 p F+2\left(m^2+4 p^2\right) p^3 T -F m^2 p_0^2 \left(9 m^2+4 p^2\right) \eta _p}{2 p^3 p_0^2}.
\end{eqnarray}
$c_\text{pol}^{\text{prob}\Sigma}$ is polarization effect contributed from the redistribution of the probe fermion in the shear flow. It is defined as the corresponding parts in \eqref{CA} by
\begin{eqnarray}
c_\text{pol}^{\text{prob}\Sigma}=16N_fe^4(2\pi)^3\int_{K,Q,K',P'}\frac{1}{(Q^2)^2}
\bigg\{
\frac{(J_1-m^2J_6)}{2P\cdot uP_\perp^2}\frac{2T_{3p}^{\text{prob}}}{p}
+\frac{(J_2+m^2J_4)}{2K\cdot uP_\perp^2}\frac{2T_{3k}^{\text{prob}}}{k}
-\frac{m^2J_5}{2K\cdot uP_\perp^2}\frac{T_{2k}^{\text{prob}}}{k}
\bigg\}\bar{f}_k \bar{f}_{k'}{f}_{p'}f_p.\nonumber\\
\end{eqnarray}
After taking the phase space integral, it becomes
\begin{eqnarray}
c_\text{pol}^{\text{prob}\Sigma}=c_f F_p+c_f^{(1)} F_p^{(1)}+c_f^{(2)} F_p^{(2)}+c_h H_p+c_h^{(1)} H_p^{(1)},
 \end{eqnarray}
with the coefficients giving by 
\begin{eqnarray}
 c_f&=&c_A\Big(\frac{F m^2 \left(5 m^2+4 p_0^2\right) \eta _p}{2p^3}-\frac{F \left(2 m^4+7 m^2 p_0^2+2 p^4\right)}{2p^2 p_0}-\frac{2 \left(3 p^2+p_0^2\right)T}{2p_0^2}\Big),\nonumber\\
 c_f^{(1)}&=&c_A\Big(m^2 \eta _p \left(\frac{9 m^2 T+4 p^2 T}{p^3}-\frac{F p}{p_0}\right)+F p^2-\frac{T \left(9 m^6+16 m^4 p^2+9 m^2 p^4+4 p^6\right)}{p^2 p_0^3}\Big),\nonumber\\
 c_f^{(2)}&=&2c_A p T \Big(p-\frac{m^2 \eta _p}{p_0}\Big),\nonumber\\
 c_h &=&\frac{c_Am^2 \left(3 m^2 p_0^2 \eta _p \left(F p^2+2 p_0 T\right)-F p_0 p^3 \left(3 m^2+p^2\right)+2 p T \left(-3 m^4-4 m^2 p^2+p^4\right)\right)}{p^5 p_0^2},\nonumber\\
 c_h^{(1)}&=&-\frac{2c_A m^2 T \left(-3 m^2 p_0 \eta _p+3 m^2 p+p^3\right)}{p^3 p_0}.
\end{eqnarray}
And $F_p=2T_{3p}^{\text{prob}}/p$ and $H_p=T_{2p}^{\text{prob}}/p$ with $T_{3p}^{\text{prob}}$ and $T_{2p}^{\text{prob}}$ given in \eqref{T3probe} and \eqref{T2probe}, there expression are also presented here for completeness,
\begin{eqnarray}
F_p&=&-\frac{3 C_f N_f\zeta (3)  T }{p^5 p_0^2}\Big[F p \left(-19 m^2 p_0^3-5 p_0^5\right)+2 p \left(3 p^4-7 p_0^2 p^2+6p_0^4\right) T-3 m^2 p_0^2 \eta _p \left(F \left(p^2-8 p_0^2\right)+4 p_0 T\right)\Big]\nonumber\\
&& -\frac{2N_f C_f^{prob}\pi ^2 p_0}{p^7} \Big[(3p_0^3 p^3-2p_0 p^5)F+(8 p^5-18 p_0^2 p^3+6 p_0^4 p)T-m^2 p_0 \eta _p \left(3 F p_0 p^2-14 p^2 T+6 p_0^2 T\right)\Big],\nonumber\\
H_p&=&\frac{6C_f N_f \zeta (3) T }{p^5 p_0^2}\Big[(17p^2-30
   p_0^2)p p_0^3F+(9 p^4-7 p^2 p_0^2+15 p_0^4)p T+3 p_0^2 \eta _p \left(\left(p^4-9 p_0^2 p^2+10 p_0^4\right)F +\left(3 p^2-5 p_0^2\right)p_0  T\right)\Big]\nonumber\\
 && -\frac{2N_f C_f^{prob}\pi ^2}{6 p^7 p_0}  \Big[
 -\left(4 p^2+3 p_0^2 \right)p^3 p_0^3F 
+\left(4 p^6+2 p_0^2 p^4+9 p_0^4 p^2+3 p_0^6\right) p T\nonumber\\
 &&\qquad\qquad\qquad\quad +p_0 \eta _p \left(\left(-2  p^4+3 p_0^2 p^2+3 p_0^4 \right)p^2p_0F +\left(2 p^6+p_0^2 p^4-8 p_0^4 p^2-3p_0^6\right) T\right)\Big].
 \end{eqnarray}
$F_p^{(1)}=-\partial_{p_0}F_p(p_0,m)$, $F_p^{(2)}=\frac{1}{2}\partial_{p_0}^2F_p(p_0,m)$ and $H_p^{(1)}=-\partial_{p_0}H_p(p_0,m)$ can be obtained by taking $p_0$ and $m$ as variable and taking derivatives with respective to $p_0$.

In the small momentum limit $p\ll m$, the coefficients in \eqref{differential} are expanded as series of $v=p/m$ around $v=0$. The coefficients becomes 
\begin{eqnarray}
\label{coe_smallp}
c_\text{diff}^{(0)}\big|_{p\ll m}&=&-c_A4T,\nonumber\\
c_\text{diff}^{(1)}\big|_{p\ll m}&=&c_A2 T p,\nonumber\\
c_\text{diff}^{(2)}\big|_{p\ll m}&=&c_A\frac{2}{3}T p^2,\nonumber\\
c_\text{pol}\big|_{p\ll m}&=&c_A\Big(\frac{T \left(7 m-225 T \zeta (3) C_f N_f\right)}{50 p^2}+\frac{1}{2}+\Big(\frac{\pi ^2 (C_b-C_f)}{16}-C_fN_f\Big(\frac{9  \zeta (3)}{4 \pi ^2}+ \ln 2\Big)\Big) \nonumber\\
&&+\tanh
   \left(\frac{m}{2 T}\right) \left(\frac{27 T  C_f N_f\zeta (3)}{10 m}+\frac{67}{200}\right)-\frac{38789 T}{10500 m}+\frac{3087 C_f N_f  \zeta (3) T^2}{105 m^2}\Big).
 \end{eqnarray}
 In the large momentum limit $p\gg m$, the coefficients in \eqref{differential} are expanded as series of $v=p/m$ around $v=\infty$. Keeping the leading and subleading orders, the coefficients becomes 
\begin{eqnarray}
\label{coe_largep}
c_\text{diff}^{(0)}\big|_{p\gg m}&=&-c_A(p+4),\nonumber\\
c_\text{diff}^{(1)}\big|_{p\gg m}&=&-c_A(p^2-4p),\nonumber\\
c_\text{diff}^{(2)}\big|_{p\gg m}&=&c_Ap^2,\nonumber\\
c_\text{pol}\big|_{p\gg m}&=&c_A(\frac{3}{2}p+6).
 \end{eqnarray}

\end{appendix}

\bibliographystyle{unsrt}
\bibliography{static_solution}

\end{document}